\documentclass[sigconf, nonacm]{acmart}
\pdfoutput=1
\usepackage[normalem]{ulem}

\startPage{1}

\setcopyright{none}

\usepackage{booktabs}   
\usepackage[figure, linesnumbered, inoutnumbered, commentsnumbered, noend] {algorithm2e}
\usepackage{mdwmath}
\usepackage{mdwtab}
\usepackage{multirow}
\usepackage[normalem]{ulem}
\usepackage{stfloats}
\usepackage{url}
\usepackage{lipsum}  
\usepackage{color}
\usepackage{xcolor}
\usepackage{listings}
\usepackage{enumitem}
\usepackage{amsmath}
\usepackage{amssymb}
\usepackage{array}
\usepackage[font=small]{caption}

\newcolumntype{P}[1]{>{\centering}p{#1}}
\newcolumntype{M}[1]{>{\centering}m{#1}}

\makeatletter
\renewcommand{\algocf@Vline}[1]{
  \strut\par\nointerlineskip
  \algocf@push{\skiprule}
  \hbox{\bgroup\color{lightgray}\vrule\egroup%
    \vtop{\algocf@push{\skiptext}
      \vtop{\algocf@addskiptotal #1}\bgroup\color{lightgray}\Hlne\egroup}}\vskip\skiphlne
  \algocf@pop{\skiprule}
  \nointerlineskip}
\renewcommand{\algocf@Vsline}[1]{
  \strut\par\nointerlineskip
  \algocf@bblockcode%
  \algocf@push{\skiprule}
  \hbox{\bgroup\color{lightgray}\vrule\egroup
    \vtop{\algocf@push{\skiptext}
      \vtop{\algocf@addskiptotal #1}}}
  \algocf@pop{\skiprule}
  \algocf@eblockcode%
}
\SetKwProg{Fn}{Function}{}{}
\SetKwProg{uForEach}{foreach}{ do}{}

\newcommand{\pycuasm}{pyReDe}

\begin{document}

\title{RegDem: Increasing GPU Performance via Shared Memory Register Spilling}
\date{}

\author{Putt Sakdhnagool}
\affiliation{
  \institution{National Electronics and Computer Technology Center}             
  \postcode{12120}
  \country{Thailand}
}
\email{putt.sakdhnagool@nectec.or.th} 

\author{Amit Sabne}
\affiliation{
  \institution{Google Brain}            
  \state{CA}
  \country{USA}
}
\email{asabne@google.com}         

\author{Rudolf Eigenmann}
\affiliation{
  \institution{University of Delaware}            
  \state{DE}
  \country{USA}
}
\email{eigenman@udel.edu}         

\begin{abstract}
GPU utilization, measured as \emph{occupancy}, is limited by the parallel threads' combined usage of on-chip resources, such as registers and the programmer-managed shared memory. Higher resource demand means lower effective parallel thread count, and therefore lower program performance. Our investigation found that registers are often the occupancy limiters. 

The de-facto \lstinline{nvcc} compiler-based approach spills excessive registers to the off-chip memory, ignoring the shared memory and leaving the on-chip resources underutilized.  To mitigate the register demand, this paper presents a binary translation technique, called \emph{RegDem}, that spills excessive registers to the underutilized shared memory by transforming the GPU assembly code (SASS). Most GPU programs do not fully use shared memory, thus allowing RegDem to use it for register spilling. The higher occupancy achieved by RegDem outweighs the slightly higher cost of accessing shared memory instead of placing data in registers. The paper also presents a compile-time performance predictor that models instructions stalls to choose the best version from a set of program variants. Cumulatively, these techniques outperform the \lstinline{nvcc} compiler with a 9\% geometric mean, the highest observed being 18\%.

\end{abstract}


\maketitle

\sloppy
\section{Introduction}
\label{sec:introduction}

Thousands of threads can concurrently reside on GPU multiprocessors (SM). Each SM contains on-chip resources, such as registers and shared memory, used by the running threads. When the cumulative demand of the resources exceeds the resource capacity, some threads need to be suspended, decreasing multiprocessor utilization, called \emph{occupancy}~\cite{cuda-best-practice}. 


Achieving maximum occupancy is challenging due to limited on-chip resources. Among all on-chip resources, registers are the most common occupancy limiting factor. For example, NVIDIA Maxwell GPU architectures provide up to 64k 32-bit registers and 2048 resident threads per multiprocessor. Each thread can use only 32 registers if maximum occupancy is desired. Such limitation could easily be surpassed by medium-sized kernels ($\sim$100 lines of code). When registers limit occupancy, reducing just a few registers could significantly improve occupancy due to the step-function behavior of occupancy with respect to the kernel's register requirement~\cite{cuda-occupancy-calc}. 


The default GPU compiler, \lstinline{nvcc}, provides an option to perform aggressive register allocation to emit binaries with fewest spills. \lstinline{nvcc} achieves that by choosing instruction sequences that need fewer registers but are less efficient, e.g., it re-materializes~\cite{Briggs:1992} 
expressions. The excessive registers are then spilled to the thread-private, off-chip memory space called \emph{local memory}~\cite{nvcc}. While such method minimizes the overhead of local memory accesses, less efficient binaries are created. 

This paper presents an alternative scheme, termed {\em register demotion}, referred to as \emph{RegDem} hereafter. RegDem increases program occupancy by spilling to the shared memory instead. The approach performs better due to the following reasons: 1) The GPU shared memory is on-chip and software managed. Therefore, the effective latency is substantially less compared to the local memory, even with a hardware-managed cache. 2) Although shared memory is primarily intended for programmer use, 
it often has enough storage available for register spilling. We  observed that, in all our target applications, there is sufficient shared memory space for spilling. 3) As mentioned earlier, \lstinline{nvcc}, under the aggressive register allocation option, avoids spilling to local memory as much as possible, owing to its high effective latency. Instead, \lstinline{nvcc} produces slower instruction sequences, reducing overall GPU performance.
Because RegDem spills to programmer-managed shared memory, it incurrs no such slowdowns.

Prior work has proposed several approaches to reduce register pressure on GPUs~\cite{hayes-local-spill,hayes-orion,sampaio-reg-spill,xie-register-allocation,you-vector-register-alloc}. The closest related GPU register allocation algorithm~\cite{hayes-local-spill} operates on the binary generated by \lstinline{nvcc} with aggressive register allocation. This algorithm converts the spills to the faster shared memory, provided sufficient space is available. While this approach, like RegDem, takes advantage of faster memory, it cannot achieve full benefits due to reason (3) above.

RegDem could be efficiently implemented during a common register allocation process. Because the \lstinline{nvcc} compiler infrastructure is proprietary, we implemented RegDem in a custom binary translation pass. RegDem determines the count of registers to be spilled, and subsequently translates the user-provided, efficient \lstinline{nvcc}-generated binary. 
The algorithm then compacts the no longer contiguous register space, minimizing the highest-used register number. This is needed because the GPU ISA determines the register usage by this number. 
Furthermore, because register allocation and instruction scheduling are interacting compiler passes, our optimization  considers the effect on the instruction schedule and performs updates where needed. The mechanism also addresses possible register bank conflicts~\cite{maxas} and the allocation of multi-word registers. 

While RegDem uses fast memory for spilling, the benefits may not always outweigh spilling overheads. Moreover, when the number of excessive registers is small, the tradeoff between aggressive register allocation and RegDem's overheads becomes non-trivial. To overcome these difficulties, we developed a compile-time performance predictor that analyzes different code variants, using instruction stalls as performance metric. It considers the combined effect of code efficiency (number of stalls) and occupancy (resulting from the number of registers and shared memory used). We then use this predictor to choose the best code variant. 

In summary, the contributions of this paper include: 
\begin{itemize}
\item A GPU register optimization algorithm, called \emph{RegDem}, which spills excessive registers to shared memory, increasing occupancy and thus performance. 

\item A compile-time performance predictor, which chooses the best code variant from among different register allocation methods. 

\item A binary translator for GPU assembly (SASS), named 
\emph{\pycuasm}
\footnote{Will be available upon paper publication.} 
that implements the proposed techniques. 

\item An integration and evaluation of the proposed performance predictor. The predictor achieves 99\% of the performance of exhaustive search for the best optimization variant. 
\end{itemize}

Our results show that RegDem with the predictor achieve 9\% geometric mean speedup over \lstinline{nvcc} on nine different benchmarks.

\section{Background}
\label{sec:background}

This section describes the Maxwell GPU architecture.\footnote{The newer architectures are similar in principle with respect to this paper.} GPU cores are organized into streaming multiprocessors (SMs). Several threadblocks can reside in each SM. Each threadblock consists of a group of threads, called \textit{warps}. Warp represents the SIMD width on GPUs. Overall, 2048 threads can reside in an SM. Occupancy is defined as the ratio of actual threads residing in an SM to the maximum. The GPU scheduling mechanism swaps warps to hide memory latency; lower occupancy results in decreased memory latency hiding, and thereby lower performance. Threads of a given threadblock can access the on-chip programmer-managed cache, termed shared memory. 

The 64K 32-bit registers available on each SM are shared among all resident threadblocks. Thus, if the register requirement per thread is high, the cumulative requirement may exceed the  available register space, allowing fewer resident threadblocks on the SM. In this manner, the register count can limit occupancy. The achieved occupancy is a step function of the kernel's register count, resulting in occupancy cliffs even when register count changes slightly.

The registers are organized across banks, and a warp accessing registers from the same bank causes higher access latency due to bank conflicts. The \lstinline{nvcc} compiler allocates registers in a manner that reduces bank conflicts. Users can limit the number of registers used by a kernel by specifying the limiting number using \lstinline{--maxrregcount} flag.

Each SM uses a hardware-managed L1 cache. Because several threadblocks share the cache, it faces higher contention. The shared memory, on the other hand, avoids the contention. Shared memory is allocated either statically, or dynamically, which means the allocation sizes only become apparent during the GPU kernel launch. The shared memory is organized into banks; threads in a warp accessing memory in the same bank see longer latencies. It is the programmer's responsibility to avoid such access patterns.

\section{Register Demotion}
\label{sec:reg-demote}

RegDem's goal is to reduce the register usage of a GPU kernel, such that the kernel achieves a higher occupancy level. Section~\ref{sec:technical_challenges} describes the challenges involved. Sections~\ref{sec:reg-demote-alg} and ~\ref{sec:reg-relocation} describe spilling algorithms, while Section~\ref{sec:optimization} presents post-spilling optimizations.

\begin{figure}[thb!]
  \centering
  \includegraphics[width=2.8in]{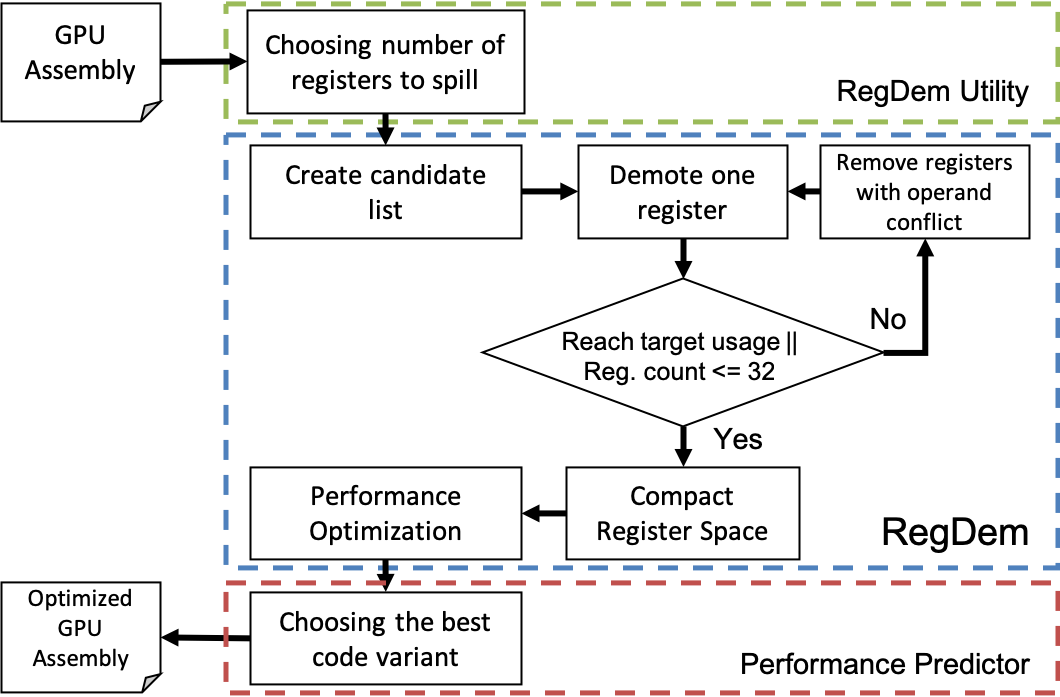}
  \caption{Register Demotion Process}
  \label{fig:register-demotion}
\end{figure}

RegDem begins with two entities: the GPU executable and a count of registers to spill. Figure~\ref{fig:register-demotion} shows overall process. RegDem contains an automatic utility that chooses different register counts to spill such that different occupancy cliffs could be achieved and the spills can fit in the available shared memory. Alternatively, the user may specify the register count to be spilled. 

Next, from the GPU executable, RegDem generates a list of candidate registers for demotion using certain selection criteria. In each iteration, one register is picked from this list and is demoted to shared memory. We will refer to the corresponding memory locations as \textbf{\emph{demoted registers}} and to the demotion code as \textbf{\emph{demoted loads and stores}}. This process repeats until the kernel reaches a desired occupancy level or the kernel's register usage falls below 32 registers, at which point further demotion offers no occupancy benefits.

\subsection{Key Challenges}
\label{sec:technical_challenges}
To realize the described approach the RegDem algorithm must address the following issues:
\begin{enumerate}[leftmargin=*]

\item{\textbf{Shared Memory Bank Conflicts:}} 
GPU shared memory consists of 32  banks. If multiple threads in a warp access different single words (32-bits) in the same bank, the accesses get serialized, increasing memory latency. The demoted loads and stores therefore must avoid  such \emph{shared memory bank conflicts}.

\item{\textbf{Operand Conflicts:}}
In GPU architectures, each shared memory access requires an explicit load/store instruction, thus needing a temporary register to access demoted registers. An \emph{operand conflict} occurs when two registers are operands of the same instruction. Demoting both registers in that situation would require two temporaries, creating additional register pressure. 

\item{\textbf{Multi-word Data Types:}}
All GPU registers are single-word, 32-bit wide. Storing multi-word data, such as double-precision floating point numbers, requires an \emph{aligned} sequence of registers in which the leading register must be even numbered. Moreover, this requirement creates \emph{register aliases}~\cite{smith-register-allocation}. For example, if R8 is being used as a double-word register, then R9 is being used implicitly, in spite of no explicit reference in the binary. 

\item{\textbf{Managing Instruction Barriers:}}
Maxwell was the first GPU ISA to introduce six different instruction barriers for synchronizing long-latency instructions~\cite{maxas}. Demoted loads and stores must therefore set barriers to signal their completion. Careful handling of these barriers, avoiding interference with existing synchronization, is essential for correctness. Furthermore, a good choice of the barriers is important for reducing stalls.

\item{\textbf{Using contiguous register numbers:}} The architecture determines the register count of a GPU kernel by the highest register number being used. E.g., if a kernel uses register \lstinline{R15}  but not \lstinline{R0} to \lstinline{R14}, the GPU will still reserve 16 registers per thread. The spilled registers will 
create \emph{gaps} in the register space, which will need to be compacted.

\item{\textbf{Register Bank Conflicts:}} The GPU register file is split into banks. If an instruction tries to access two or more registers from the same bank, the accesses will be serialized, due to a \emph{register bank conflict}~\cite{maxas}. We found  that such conflicts can increase  computation time by as much as 12\%. Temporary registers allocated by RegDem, and during compaction should avoid register conflicts.

\item{\textbf{Instruction Scheduling:}}
Because demoted loads and stores are inserted in a previously schedule-optimized executable, additional instruction scheduling opportunity presents itself. 

\item{\textbf{Choosing Candidates for Register Demotion:}}
The choice of registers to be demoted can greatly impact the performance, due to the cost of memory accesses and potential operand conflicts. 
\end{enumerate}


\subsection{Register Spilling to Shared Memory}
\label{sec:reg-demote-alg}

\begin{figure}[tb!]
  \centering
  \includegraphics[width=2.7in]{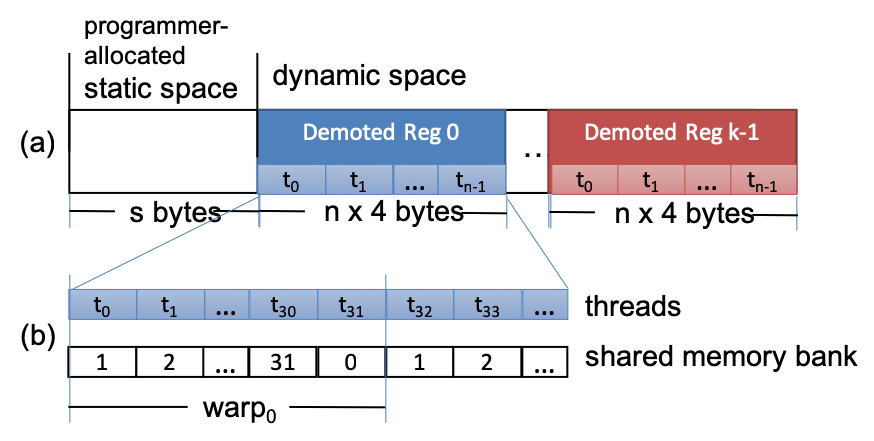}
  \caption{Demoted Registers in Shared Memory: (a) Shared memory allocation 
  with $k$ demoted registers for a kernel with $n$-threads per thread block and $s$ bytes of 
  statically allocated shared memory. 
  $t_i$ denotes the $i$-th thread in a thread block. (b) Mapping between threads inside a warp and shared memory banks when the last byte of the static allocation resides in the 0-th memory bank.
  }
  \vspace{-0.2cm}
  \label{fig:shared-memory}
\end{figure}

\paragraph{\textbf{Shared Memory Allocation for Spilling:}}
\label{sec:shared-alloc}
The allocated shared memory for spilling must avoid shared memory bank conflicts. RegDem's mechanism to that end is captured in Figure~\ref{fig:shared-memory}a. Each demoted register is assigned a contiguous space in memory, 
where successive threads own consecutive words. For a kernel with $n$ threads per thread block, each demoted register allocates $n\times4$ bytes of consecutive memory. The shared memory location of the $r$-th demoted register of the $t$-th thread of the thread block can be computed as 
\begin{equation}
\label{eq:reg_location}
location = \underbrace{(t \times 4)}_\textrm{base address} + \underbrace{s + (r \times n \times 4)}_\textrm{register offset}
\end{equation}
where $s$ is the static allocation size of the shared memory used by the kernel, rounded up to the nearest multiple of 4 bytes (shared memory bank alignment). Note that all values are known at compile-time except $t$, which is known at runtime. Because contiguous 32-bit shared memory values are placed in different banks, the above organization guarantees that all threads in a warp will access different shared memory banks as shown in Figure~\ref{fig:shared-memory}b. Our implementation dynamically allocates memory for demoted registers to separate user and demotion-allocated memory spaces. 

\paragraph{\textbf{Shared Memory Access Mechanism:}}
GPUs use base-plus-offset addressing mode for shared memory accesses.Load (\lstinline{LDS}) and store (\lstinline{STS}) instructions have the following format.
\begin{lstlisting}[frame=none]
            LDS RV, [RA+offset];
            STS [RA+offset], RV;
\end{lstlisting}
\lstinline{RA} is a register that contains the base address of a shared memory location and \lstinline{offset} is an immediate value for indexing. \lstinline{RV} denotes a register that contains the value of the target shared memory location. Both load and store instructions require barriers for synchronization. For a load instruction, read and write barriers are required to prevent \lstinline{RA} and \lstinline{RV} from write-after-read and read-after-write hazards, respectively. A store instruction only requires a read barrier to avoid a write-after-read hazard.  

As seen above, accessing a demoted register requires two additional registers. First register holds the value of the \emph{base address} from equation~(\ref{eq:reg_location}). We will refer to this register as \lstinline{RDA} (short for demoted base address register). We use dynamic addressing for demoted register access because the value of $t$ is only known at runtime. The second register is used to hold the value of the demoted register. This register will be referred to as \lstinline{RDV} (short for demoted value register). Therefore, at least two registers must be added to the program. More than one value registers can be used so that multiple demoted registers can be accessed simultaneously. However, doing so will increase register pressure. The RegDem algorithm therefore limits the number of value registers to one. Section~\ref{sec:optimization} will describe scenarios where more value registers can be employed. Note that for multi-word data, the value register count is increased based upon the data width.

\paragraph{\textbf{Algorithm:}}
RegDem's spilling algorithm renames the register to be demoted with \lstinline{RDV} and places the associated demoted load/store instructions next to the demoted register's accesses. Most importantly, the algorithm updates the instruction schedule by inserting barriers on the demoted loads/stores, which impacts both performance and correctness. Figure~\ref{alg:reg_demote} shows a simplified version of the algorithm, which caters to single-word registers only.

\begin{figure}[th!] 
  \vspace{-0.5cm}
  \SetInd{0.5em}{0.5em}
  \begin{algorithm}[H]
  \small
\DontPrintSemicolon
\textbf{Input:} Program $p$, Size of thread block $n$\;
  \textit{candidate\_list} $\leftarrow$ CreateDemotionCandidate($p$)\;
  \While{p.reg\_count $>$ target\_usage AND p.reg\_count $>$ 32 }{
  $r$ $\leftarrow$ dequeue(\textit{candidate\_list}) \;
  \textit{offset} $\leftarrow$ \textit{demoted\_reg\_count} $\times$ $n$ $\times$ 4 + \textit{p.shared\_size}\;
  \ForEach{instruction inst $\in$ p}{
  	\uIf{inst is \textrm{jump} OR \textrm{label}}{
    	ResetBarrierTracker(\textit{tracker})
    }
    \If{r $\in$ inst.regs}{
    	rename $r$ in \textit{inst} with RDV\;
    	\If{r \textrm{is destination register}}{
            \textit{inst}$_{\textit{sts}}$ $\leftarrow$ ``STS [RDA+offset], RDV;"\;
            \uIf{inst is high-latency instruction AND $\neg$ inst.write\_barrier}{
            	\textit{inst.write\_barrier} $\leftarrow$ GetBarrier(\textit{tracker})\;		
            }
            \textit{inst}$_{\textit{sts}}$.Wait(\textit{inst.write\_barrier})\;
            \textit{inst}$_{\textit{sts}}$\textit{.read\_barrier} $\leftarrow$ GetBarrier(\textit{tracker})\;
            Add \textit{inst}$_{\textit{sts}}$ after \textit{inst}\;
            \textit{inst}$_{\textit{next}}$ $\leftarrow$ NextInstruction($p$, \textit{inst}$_{\textit{sts}}$)\;
            \textit{inst}$_{\textit{next}}$.Wait(\textit{inst}$_{\textit{sts}}$\textit{.read\_barrier})\;
        }
        \If{r is operand}{
            \textit{inst}$_{\textit{lds}}$ $\leftarrow$ ``LDS RDV, [RDA+offset];"\;
			\textit{inst}$_{\textit{lds}}$\textit{.read\_barrier} $\leftarrow$ GetBarrier(\textit{tracker})\;
            \textit{inst}$_{\textit{lds}}$\textit{.write\_barrier} $\leftarrow$ GetBarrier(\textit{tracker})\;
            \textit{inst}.Wait(\textit{inst}$_{\textit{lds}}$\textit{.read\_barrier})\;
            \textit{inst}.Wait(\textit{inst}$_{\textit{lds}}$\textit{.write\_barrier})\;
            Add \textit{inst}$_{\textit{dem}}$ before \textit{inst}\;
            \textit{inst}$_{\textit{prev}}$ $\leftarrow$ PrevInstruction($p$, \textit{inst}$_{\textit{lds}}$)\;
            \If{\textit{inst}$_{\textit{prev}}$ is demoted store}
            {
            	\textit{inst}$_{\textit{lds}}$.waitBarrier(\textit{inst}$_{\textit{prev}}$\textit{.read\_barrier})
            }
        }
    }
    UpdateBarrierTracker(\textit{tracker}, \textit{inst})\; 
  }
  RemoveOperandConflict(\textit{candidate\_list}, $r$)\;
  }
 \Fn{UpdateBarrierTracker(\textit{tracker}, \textit{inst})}{
	\uIf{inst.read\_barrier}{
    	\textit{tracker}[\textit{inst.read\_barrier}]\textit{.inst} $\leftarrow$ \textit{inst}\;
        \textit{tracker}[\textit{inst.read\_barrier}]\textit{.stall} $\leftarrow$ 0
    }
    \uIf{inst.write\_barrier}{
    	\textit{tracker}[\textit{inst.write\_barrier}]\textit{.inst} $\leftarrow$ \textit{inst}\;
        \textit{tracker}[\textit{inst.write\_barrier}]\textit{.stall} $\leftarrow$ 0\;
    }
    \uForEach{barrier b $\in$ \textit{tracker}}{
    	\textit{tracker}[\textit{b}]\textit{.stall} $\leftarrow$ \textit{tracker}[\textit{b}]\textit{.stall} + \textit{inst.stall}
    }
    \uForEach{barrier b $\in$ \textit{inst}.wait}{
    	\textit{tracker}[\textit{b}] $\leftarrow$ NULL
    }
}

 \Fn{GetBarrier(tracker)}{
	\textit{min}$_{\textit{barrier}}$ $\leftarrow$ NULL; \textit{min}$_{\textit{stall}}$ $\leftarrow$ GL\_MEM\_STALL + 1\;
	\uForEach{barrier b $\in$ \textit{tracker}}{
    	\lIf{\textit{tracker}[\textit{b}] == NULL}{\Return \textit{b}}
        \uIf{\textit{tracker}[\textit{b}]\textit{.inst} is global memory inst}{
				\textit{stall} $\leftarrow$ GL\_MEM\_STALL - \textit{tracker}[\textit{b}]\textit{.stall}\;
			}
     \uElseIf{tracker[b].inst is shared memory inst}{
			\textit{stall} $\leftarrow$ SH\_MEM\_STALL - \textit{tracker}[\textit{b}]\textit{.stall}\;
        }
        	\uIf{min$_{\textit{stall}}$ $>$ stall}{
              	\textit{min}$_{\textit{barrier}}$ $\leftarrow$ \textit{b}; \textit{min}$_{\textit{stall}}$ $\leftarrow$ \textit{stall}\;
            }
    }
    \Return \textit{min}$_{\textit{barrier}}$
}
\vspace{-0.5cm}
\end{algorithm}
\caption{ RegDem Algorithm: Each iteration of the main algorithm (lines 3--31) removes one register from the program. The \lstinline{UpdateBarrierTracker} function keeps track of barrier usage. The \lstinline{GetBarrier} function uses this information to select the barrier that will cause the least stalls. }
\label{alg:reg_demote}
\vspace{-0.3cm}
\end{figure}

The main algorithm (lines 3--31) spills one register at a time until the register usage reaches the target level or falls below 32, where it no longer limits occupancy. In each iteration, a register $r$ is picked from the candidate list and its shared memory location is computed (lines 4--5). Next, the algorithm  searches and replaces $r$ with the \lstinline{RDV} register (line 10). A shared memory instruction for accessing/updating the demoted register value is then inserted (lines 11--29). After all occurrences of $r$ are replaced, the algorithm removes all candidates that have operand conflicts with $r$ (line 31). 

For write accesses (lines 11--19), a store instruction (\textit{inst}$_{\textit{sts}}$) is placed immediately after the write instruction (\textit{inst}) to update the demoted register's value in shared memory. The algorithm ensures that (1) updating \lstinline{RDV} has completed before storing its value to the demoted register and (2) \lstinline{RDV} is not rewritten before writing its value to the demoted register has completed. Similarly, for read accesses (lines 20--29), a load instruction (\textit{inst}$_{\textit{lds}}$) is placed before the instruction that accesses the demoted register (\textit{inst}). The algorithm inserts barriers, ensuring that the shared memory load is completed before \lstinline{RDV} is used by the next instruction. If the instruction before the demoted load (\textit{inst}$_{\textit{prev}}$) is a demoted store, the algorithm ensures that \lstinline{RDV} is free before the demoted register is loaded.

Note from above that two barriers are used by each demoted load/store instruction to synchronize with other instructions. The barriers are limited in count; only six barriers exist on Maxwell and Pascal architectures. Therefore, if the barrier placed on a demoted load/store instruction was already occupied by a different instruction, additional stalls are introduced. A poor choice of a barrier may result in a wait of as many as 200 cycles, if that barrier is busy. To minimize the synchronization overhead, the RegDem algorithm presents a \emph{barrier tracker} to monitor barrier usage. The tracker records the last instruction that will set the barrier and estimates the number of cycles passed since the setting instruction was executed as shown in the \lstinline{UpdateBarrierTracker()} function. The algorithm obtains a new barrier through the \lstinline{GetBarrier()} function. The function returns a free barrier if available, or else it returns the barrier that generates minimum stalls, which is determined by the type of setting instruction and the number of cycles passed since that instruction. Estimating the cycle count accurately is crucial to reducing the stall count. A key observation helps us estimate the stall count statically: GPU architectures require that barriers are cleared before jump instructions, and hence can only span basic blocks. The barrier tracker therefore can mark barriers that are assigned before a jump instruction to be definitely available after the jump. For straight line code, the tracker estimates cycle counts per instruction based on instruction latencies. For example, 
our current implementation sets the latency of device memory access (\lstinline{GL_MEM_STALL}) to 200 cycles\cite{cuda-programming-guide}, and the latency of shared memory access (\lstinline{SH_MEM_STALL}) to 24 cycles, based on the stalls caused by register read-after-write dependencies~\cite{cuda-best-practice}.

\paragraph{\textbf{Extension for Multi-word Data:}}
We can extend the algorithm to demote registers containing multi-word data. Recall that multi-word data requires an aligned series of registers. To accommodate this requirement, the algorithm chooses \lstinline{RDV} to be even-numbered and adds extra registers for padding if needed. In our implementation, each register in the series is treated individually and uses the same allocation scheme as single-word registers. This allocation scheme allows each register to be accessed separately while avoiding bank conflicts. When accessing multi-word demoted registers, multiple load/store instructions are inserted. 

\subsection{Register Compaction}
\label{sec:reg-relocation} 

\begin{figure}[tb!]
  \centering
  \includegraphics[width=2.6in]{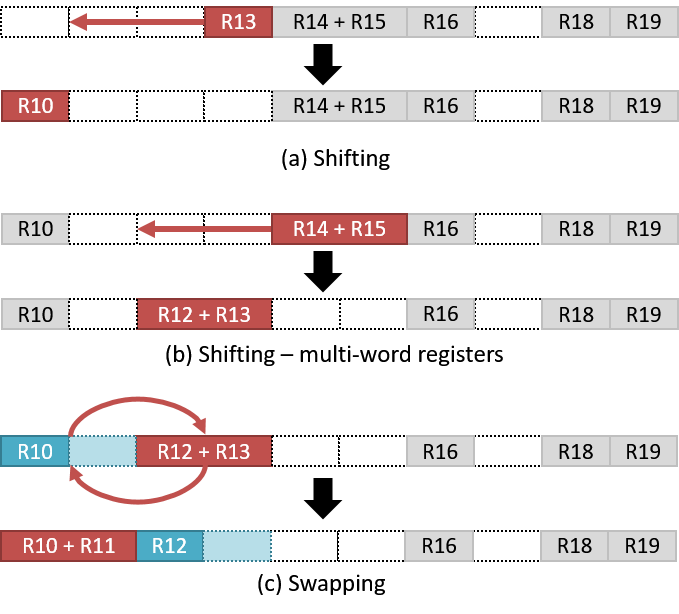}
  \caption{Register Relocation Space and Register Compaction: A multi-word register is represented by a series of registers connected using `$+$' signs. (a) Single-word register shifting. (b) Double-word register shifting. The register cannot be moved to the first gap because of alignment restriction. (c) A double-word register is swapped with a swapping window of size 2 (marked in cyan) containing a single-word register and a gap.} 
  \label{fig:reg-relocate}
\end{figure}

Recall that the last physical register number present in the code determines the register usage of the kernel. Demoted registers may still count as used until this number is reduced. Compaction achieves that effect. Our algorithm uses a data structure called \emph{relocation space}, which utilizes an array for performing virtual register movement. Figure~\ref{fig:reg-relocate} shows this data structure. Each slot represents one physical register present in the program. Multi-word registers occupy multiple slots based on their size and are represented as single registers. This representation prevents the algorithm from breaking register aliases when compressing the register space. A register gap is represented by an empty slot.  

The algorithm pushes the gaps toward the end of the register space by using two operations, \emph{shifting} and \emph{swapping}. Figure~\ref{fig:reg-relocate} shows an example of the relocation space and its operations. Shifting moves the next available register to fill the gaps (Figure~\ref{fig:reg-relocate}a). Occasionally, the shifting operation cannot move the multi-word register to the available gap due to register alignment (Figure~\ref{fig:reg-relocate}b). In this situation, swapping applied after the shifting operation moves the register to the closest possible gap. The swapping operation exchanges the multi-word register with the registers in the \emph{swapping window} (Figure~\ref{fig:reg-relocate}c). The swapping window starts from the location of the multi-word register and grows toward the lower-numbered location. The size of swapping windows is determined by the size of the multi-word registers. 

  
  

\subsection{Post-Spilling Optimizations}
\label{sec:optimization}

This section discusses optimization opportunities presented after applying RegDem's shared memory spilling.

\subsubsection{\textbf{Avoiding Register Bank Conflicts}}
\label{sec:reg-bank-conflict}
When register operands in the same instruction belong to the same register bank, their accesses get serialized. Because \lstinline{RDV} is chosen before RegDem begins, bank conflicts may get introduced during the demotion process.

The first strategy is to choose \lstinline{RDV} from the bank that generates the least number of conflicts. This is achieved with a small addition to the presented algorithm, keeping track of the conflicts caused by \lstinline{RDV}. Register bank conflict avoidance is also added to the compaction algorithm (Section~\ref{sec:reg-relocation}). The mechanism searches for registers from the same bank to fill gaps. Swapping is performed within a window of size equal to the number banks in the register file, i.e., four. This modification can lose efficiency when an even-numbered gap is filled by a single-word register, leaving a small gap. We revert to the original algorithm in that case since reducing register count is the top priority.

\subsubsection{\textbf{Performance Enhancements}}
\label{sec:optimize}
RegDem adds load/store instructions conservatively for lack of global analysis. The following optimization passes improve the code in these situations: 


\paragraph{\textbf{Eliminating redundant demote code:}}
This pass reduces demoted register access overhead by tracking the recent value of the \lstinline{RDV} register and eliminating subsequent loads of the same, if the value is still alive. For example, if two consecutive instructions read the same demoted register, a demoted load will be placed before each of them. This optimization will remove the second load. Similarly, demoted stores are removed if their target demoted register will be updated again in the near future. 

% ...


\paragraph{\textbf{Updating instruction schedule:}}
This pass reschedules demoted loads and stores to reduce instruction synchronization overhead. Demoted loads are hoisted as early as possible, updating the associated instruction barriers. The analysis also removes barriers from demoted stores if the \lstinline{RDV} register will not be updated before the memory store completes.

\paragraph{\textbf{Substituting Value Register:}} 
To keep register pressure low, only one \lstinline{RDV} register is reserved for accessing demoted registers. Hence, only one demoted register can be in use at any given time, restricting the window within which demoted loads/stores can be hoisted. To enlarge this window, the optimization analyzes register liveness in each basic block, identifying other free registers as local temporaries. Then, the optimization substitutes \lstinline{RDV} inside the block with these temporaries, allowing multiple demoted registers to be in use simultaneously without increased register pressure.

\subsubsection{\textbf{Choosing Candidate Registers for Demotion}}
\label{sec:choosing-reg}

We use three strategies for choosing candidate registers. Each strategy estimates register access counts, and candidates are chosen in  ascending order of the access count. The first strategy makes a simple pass through the assembly code and counts the number of static accesses for each register. The second strategy traverses the CFG to count register accesses of each basic block. For basic blocks inside a loop, the access count is multiplied by a generic value of 10. The third strategy takes operand conflicts into consideration. It chooses candidates in ascending order of their operand conflicts. The performance predictor described in the next section (Section~\ref{sec:offline-tuning}) chooses from among these three strategies.

\section{Compile-time Performance Predictor}
\label{sec:offline-tuning}

In some corner cases, the RegDem benefits may not  outweigh the spilling overheads. When the spill count is small, the tradeoff between aggressive register allocation and overheads of RegDem becomes non-trivial. With the optimizations described in Section~\ref{sec:optimization}, the difficulty of choosing the best performing code variant increases further. 

To make that decision, we developed a compile-time performance predictor that analyzes GPU binaries and selects the best code, also considering non-RegDem variants. The predictor approximates program performance using instruction stalls as performance metric. The predictor considers both explicit stalls, presented in instruction annotations~\cite{maxas,volta-architecture}, and implicit overheads from (i)~memory accesses latencies, (ii)~variations in instruction throughput across different instruction types, and (iii)~loop and function constructs. Figure~\ref{alg:perf-estimation} shows the performance predictor algorithm. It performs three main steps:

Step one (lines 3--22) traverses through the program control flow graph (CFG) and estimates the stall cycles in each basic block. The algorithm collects stalls generated by each instruction in the block and adjusts these stalls based on instruction throughput and memory accesses. GPU instructions could have different throughput based on available resources, e.g., Maxwell GPUs have 128 FP32 and 4 FP64 cores. Instructions with less resources would experience more stalls due to higher contention. The predictor factors in occupancy and instruction throughput,  using the following equation:
\begin{equation}
\small
\label{eq:inst-stall}
stall = inst_{\textit{stall}} \times occupancy \times \frac{\textit{MAX\_THROUGHPUT}}{\textit{inst}_{\textit{throughput}}}
\end{equation}
$inst_{\textit{stall}}$ denotes the stall cycles per the instruction annotation.
The term $\frac{\textit{MAX\_THROUGHPUT}}{inst_{\textit{throughput}}}$ estimates the contention of lower-throughput instructions, where $\textit{MAX\_THROUGHPUT}$ denotes the maximum instruction throughput and $\textit{inst}_{\textit{throughput}}$ denotes the throughput of the instruction. For Maxwell GPUs, the value of $\textit{MAX\_THROUGHPUT}$ is 128 instructions/cycle. $occupancy$ is used for estimating the number of threads waiting for the resource. 

\begin{figure}[t!] 
  \vspace{-0.4cm}
  \SetInd{0.5em}{0.5em}
  \begin{algorithm}[H]
  \small
  \DontPrintSemicolon
  \KwIn{Program $p$, Program CFG \textit{cfg}} 
  \KwOut{Estimated stall cycle \textit{stall}$_{\textit{count}}$}
    
  \For{block $\in$ cfg}{
  	\textit{block.stall} $\longleftarrow$ 0\;
    \For{inst $\in$ block.instructions}{
        
        \textit{inst.stall} $\leftarrow$ \textit{inst.stall} $\times$ \textit{p.occupancy} $\times$ $\frac{\textrm{MAX\_THROUGHPUT}}{\textit{inst.throughput}}$ \;	

    	\If{inst.read\_barrier}{
        	\textit{tracker}[\textit{inst.read\_barrier}]\textit{.inst} $\longleftarrow$ \textit{inst}\;
            \textit{tracker}[\textit{inst.read\_barrier}]\textit{.stall} $\longleftarrow$ 0\;
        }
        \If{inst.write\_barrier}{
        	\textit{tracker}[\textit{inst.write\_barrier}]\textit{.inst} $\longleftarrow$ \textit{inst}\;
            \textit{tracker}[\textit{inst.write\_barrier}]\textit{.stall} $\longleftarrow$ 0\;
        }
        
        \For{w $\in$ inst.wait\_barriers}{
            \uIf{tracker[w].inst is global access}
            {
                \If{tracker[w].stall $<$ GL\_MEM\_STALL}
                {
        			\textit{block.stall} $\longleftarrow$ \textit{block.stall} + GL\_MEM\_STALL $-$ \textit{tracker}[\textit{w}]\textit{.stall} \;        	
                }
            }
            \ElseIf{tracker[w].inst is shared access}
            {
                \If{tracker[w].stall $<$ SH\_MEM\_STALL}
                {
        			\textit{block.stall} $\longleftarrow$ \textit{block.stall} + SH\_MEM\_STALL $-$ \textit{tracker}[\textit{w}]\textit{.stall} \;        	
                }
            }
        }
        \For{bar $\in$ barriers}{
        	\textit{tracker}[\textit{bar}]\textit{.stall} $\longleftarrow$ \textit{tracker}[\textit{bar}]\textit{.stall} + \textit{inst.stall}
        }

        \textit{block.stall} $\longleftarrow$ \textit{block.stall} + \textit{inst.stall}\;
    }
  }
  \For{block $\in$ cfg in breath-first order}{
  	\For{edge $\in$ block.edge}{
    	\If{edge is backward}{
        	\textit{loop} $\longleftarrow$ GetLoop(\textit{block}, \textit{edge})\;
        	\For{ b $\in$ loop.blocks}
            {
            	 \textit{b.stall} $\leftarrow$ \textit{b.stall} $\times$ LOOP\_FACTOR \;
            }
        }
    }
  }
  \textit{stall}$_{\textit{count}}$ $\longleftarrow$ 0\;
  \For{block $\in$ cfg in breath-first order}{
  	\textit{stall}$_{\textit{count}}$ $\longleftarrow$ \textit{stall}$_{\textit{count}}$ + \textit{block.stall}\;
  }
\end{algorithm}
\vspace{-0.4cm}
\caption{Performance Estimation Algorithm}
\label{alg:perf-estimation}
\end{figure}

Memory access stalls are estimated by tracking the time 
between barrier set and register use, then taking the maximum of this time and the memory latency. The algorithm uses the barrier tracker and memory latency described in Section~\ref{sec:reg-demote} for analysis. 

Step two (lines 23--28) updates the stall count of every basic block inside a loop. The block stall count is multiplied by a generic $\textit{LOOP\_FACTOR}$; the current implementation sets this value to 10, which is a plausible static estimate for a value that would need dynamic analysis. This method weighs loops higher than straight-line code. 

Step three (lines 29--31) estimates the overall stalls 
by summing stall cycles of all basic blocks. This approximation is needed because branch decisions are not known statically.
Furthermore, both branch targets must be considered because the GPU SIMD approach results in serial execution of branch taken/not taken paths if even a single thread executes the respective path. The algorithm proceeds interprocedurally,  estimating the CFGs of inner functions first. 

To compare code variants generated by the different register allocation methods, the predictor considers stalls as well as program occupancy. Improving occupancy normally yields  diminishing returns, and degrades performance at worst~\cite{volkov-occupancy}. The predictor reflects this behavior in the estimated execution time using the following equation to adjust the result $stall_{\textit{count}}$ from Figure~\ref{alg:perf-estimation}.
\begin{equation}
\small
\label{eq:stall-adjust}
stall_{\textit{program}} = \frac{f(occupancy)}{f(occupancy_{\textit{max}})} \times stall_{\textit{count}}
\end{equation}
$stall_{\textit{program}}$ represents an estimated execution time of the code
variant in stall cycles. The term $\frac{f(occupancy)}{f(occupancy_{\textit{max}})}$ computes the slowdown caused by the lower occupancy, where $occupancy_{\textit{max}}$ is the maximum occupancy across code variants and $f(x)$ is a function used for estimating the execution time at $x$\% occupancy. $f(x)$ was determined empirically, using compute-intensive microbenchmarks across various thread block sizes. The occupancy of the microbenchmarks is controlled by modifying register usage, measuring only the impact of occupancy on performance. The predictor uses theoretical occupancy in the computation, which can be computed from the thread block size of the user input~\cite{cuda-occupancy-calc}.

\section{Evaluation}
\label{sec:eval}

This section evaluates the presented RegDem technique and performance predictor. We compare the performance of RegDem  with the default \lstinline{nvcc} code generation and the closest research alternative~\cite{hayes-local-spill}. Next, we measure the impact of the optimization options. Lastly, we evaluate the added gain by the performance predictor.

\begin{table*}[tb!]
    \vspace{6pt}
	\caption{Details of Benchmark Kernels used in Performance Evaluation. }
	\label{tab:benchmarks-detail}
    \footnotesize
	\centering
	\begin{tabular}{| l | l | m{1.5cm} | r | r | r | r | r | r | r | r | r |}
		\hline
		\multirow{2}{0.5cm}{Bench-mark} & \multirow{2}{*}{Kernel} & \multirow{2}{*}{Input} & \multirow{2}{0.8cm}{\centering \#Thread \newline blocks} & \multirow{2}{0.8cm}{\centering Threads \newline / block} & \multirow{2}{0.8cm}{\centering Shared memory} & \multicolumn{2}{m{1.5cm}|}{\centering \# Registers Used} & \multicolumn{2}{m{1.8cm}|}{\centering \# Registers Spilled\footnotemark[1]} & \multicolumn{2}{m{1.5cm}|}{\centering Achieved Occupancy\footnotemark[2]} \\ \cline{7-12}
        & & & & & & orig & target & nvcc & RegDem & orig & RegDem\\ \hline
		\hline
        cfd & cuda\_compute\_flux & fvcorr.193K & 1008 & 192 & 0B & 68 & 56 & 10 & 14 & 0.35 & 0.54\\ \hline
        qtc & QTC\_device & 8192 points & 1538 & 64 & 512B & 55 & 48 & 8 & 10 & 0.51 & 0.57 \\ \hline
        md5hash & FindKeyWithDigest\_Kernel & 1680M keys& 93790 & 256 & 0B & 33 & 32 & 0 & 3 & 0.70 & 0.94 \\ \hline
        md & compute\_lj\_force & 73728 atoms& 228 & 256 & 0B & 34 & 32 & 1 & 5 & 0.75 & 0.83 \\ \hline
        gaussian & d\_recursiveGaussian\_rgba & 32K$\times$10K px & 500 & 64 & 0B & 43 & 40 & 1 & 5 & 0.58 & 0.62\\ \hline
        conv & convolutionColumnsKernel & 4K$\times$4K px & 16384 & 128 & 0B & 35 & 32 & 0 & 5 & 0.73 & 0.98 \\ \hline
        nn & nearest\_neighbor\_search & 200K 7d pts & 1024 & 192 & 1.52KB & 35 & 32 & 0 & 5 & 0.55 & 0.72\\ \hline
        pc & compute\_correlation & 200K 7d pts & 1024 & 256 & 2.03KB & 36 & 32 & 2 & 6 & 0.54 & 0.72 \\ \hline
        vp& search\_kernel & 200K 7d pts & 2048 & 256 & 2.03KB& 34 & 32 & 0 & 4 & 0.52 & 0.68  \\ \hline
	\end{tabular} \\
    \raggedright{\footnotemark[1] Number of registers spilled / demoted by \lstinline{nvcc} and RegDem when restricting register usage to the specified target.
    }\\
    \raggedright{\footnotemark[2] Achieved occupancy of the kernel measured by \lstinline{nvprof} profiler before (\emph{orig}) and after RegDem.}
\end{table*}

\begin{table}[tb!]
	\caption{Benchmark Description}
	\label{tab:benchmarks}
    \footnotesize
	\centering
	\begin{tabular}{| m{1.6cm} | l | m{4.5cm} |}
		\hline
		Benchmark suite & Benchmark & Description \\ \hline 
		\hline
        Rodinia~\cite{rodinia-benchmark} & cfd & An unstructured grid solver for three-dimensional Euler equations. \\ \hline
        \multirow{3}{1.3cm}{SHOC~\cite{shoc-benchmark}} & qtc & A quality threshold clustering algorithm\\ \cline{2-3}
        & md5hash & A brute force search to find a key with a MD5 digest.\\ \cline{2-3}
        & md & An N-body computation computing the Lennard-Jones potential.\\ \hline
        \multirow{2}{1.6cm}{CUDA Toolkit-Imaging~\cite{cuda-conv, cuda-toolkit}} & gaussian & A Gaussian blur using Deriche's recursive method.\\ \cline{2-3}
        & conv & A separable convolution filter for 2D image. \\ \hline
        \multirow{3}{1.3cm}{FSM~\cite{fsm-benchmark}} & nn & Nearest neighbors search of the input points in the metric space using kd-tree. \\ \cline{2-3}
        & pc & Computing two-point correlation of each points in the input data. \\ \cline{2-3}
        & vp & Nearest neighbors search of the input points in the metric space using vantage point trees.\\ \hline
	\end{tabular}
\end{table}

\subsection{Experimental Settings}

We evaluated our techniques on an Ubuntu Linux 14.04.3 system with a quad-core Intel Core i7-6700K processor running at 4.00 GHz, 16 GB of main memory, and a Maxwell-based GeForce GTX Titan~X GPU with 12 GB device memory. 

Recall that registers can limit occupancy if and only if a kernel requires more than 32 registers per thread. Only applications where register pressure limits occupancy will benefit from RegDem. This is the case in nine applications of the four benchmark suites in Table~\ref{tab:benchmarks}. RegDem has no effect on the other
applications.

The baseline versions for all benchmarks are created with \lstinline{nvcc} and the benchmark-provided compiler flags. The optimized versions use the techniques described in this paper. RegDem extracts assembly code from a .cubin file, performs the optimizations, and regenerates assembly code. The \lstinline{MaxAs} tool~\cite{maxas} then inserts the optimized code into the original .cubin file. We used \lstinline{nvcc} version 6.5, the latest version supported by \lstinline{MaxAs}. NVIDIA's \lstinline{nvprof} profiler was used to measure the average execution time of the kernels across five runs.

\subsection{Achieved Occupancy}

Table~\ref{tab:benchmarks-detail} shows achieved occupancy of the benchmarks before (\emph{orig}) and after RegDem. On average, RegDem improves occupancy by 27\%. Benchmarks with larger thread block size could see higher improvements owing to the step-function nature of occupancy. 

\subsection{Code Variants used in Performance Evaluations}
We consider four code variants in addition to RegDem, shown in Table~\ref{tab:code-config}. The \emph{nvcc} version represents the \emph{baseline} performance. It is compiled with \lstinline{nvcc} and the default compiler flags provided by the benchmarks. 

The \emph{local} variant uses \lstinline{nvcc} with \lstinline{--maxrregcount} flag, forcing the compiler to use aggressive register allocation and spill excessive registers to local memory. This is the state-of-the-art method for restricting register usage. The register count is set to be the same as in RegDem.

\emph{Local-shared} and \emph{local-shared-relax} realize the technique presented by Hayes et al.~\cite{hayes-local-spill}, which converts spill code from \emph{local} to shared memory. The only difference is the target register usage. The \emph{local-shared} variant strictly follows the \cite{hayes-local-spill} implementation. The target register usage is set to 32 registers, allowing kernels to execute at maximum occupancy, and relies on tuning the thread block size to achieve the best performance. This is unlike our approach, where the number of spilled registers is the only parameter tuned. We consider this version the \emph{closest research alternative}. For fair comparison, \emph{local-shared-relax} relaxes the register usage restriction and sets it to the same value used in RegDem. We call this variant the \emph{enhanced research alternative}. 

\begin{table}[tb!]
	\caption{Code Variant Comparison}
	\label{tab:code-config}
    \footnotesize
	\centering
	\begin{tabular}{| m{3.0cm} | P{0.6cm} | P{0.6cm} | P{0.6cm} | P{0.6cm} | P{0.6cm} |}
		\hline 
		& nvcc (base\-line) \vspace{-0.4cm} & Reg\-Dem & local & local\-shared\vspace{-0.4cm} & local\-shared\-relax \vspace{-0.4cm}\\ \hline \hline
        spilled memory space & - & shared\vspace{-0.4cm} & local & shared\vspace{-0.4cm} & shared\vspace{-0.4cm} \\ \hline
        target register usage & $\star$ & $\dagger$ & $\dagger$ & 32 & $\dagger$ \\ \hline
        use \texttt{nvcc} to spill registers & - & - & \checkmark & \checkmark & \checkmark  \\ \hline
        convert local to shared mem. & - & - & - & \checkmark & \checkmark  \\ \hline
        demote reg. to shared memory & - & \checkmark & - & - & - \\ \hline  
	\end{tabular} \\
    \raggedright{\boldmath$\star$: not restricted, \boldmath$\dagger$: set to the target register usage specified in Table~\ref{tab:benchmarks-detail}.}
\end{table}

\subsection{Performance Results}
\begin{figure*}[thb!]
\centering
\includegraphics[width=4.3in]{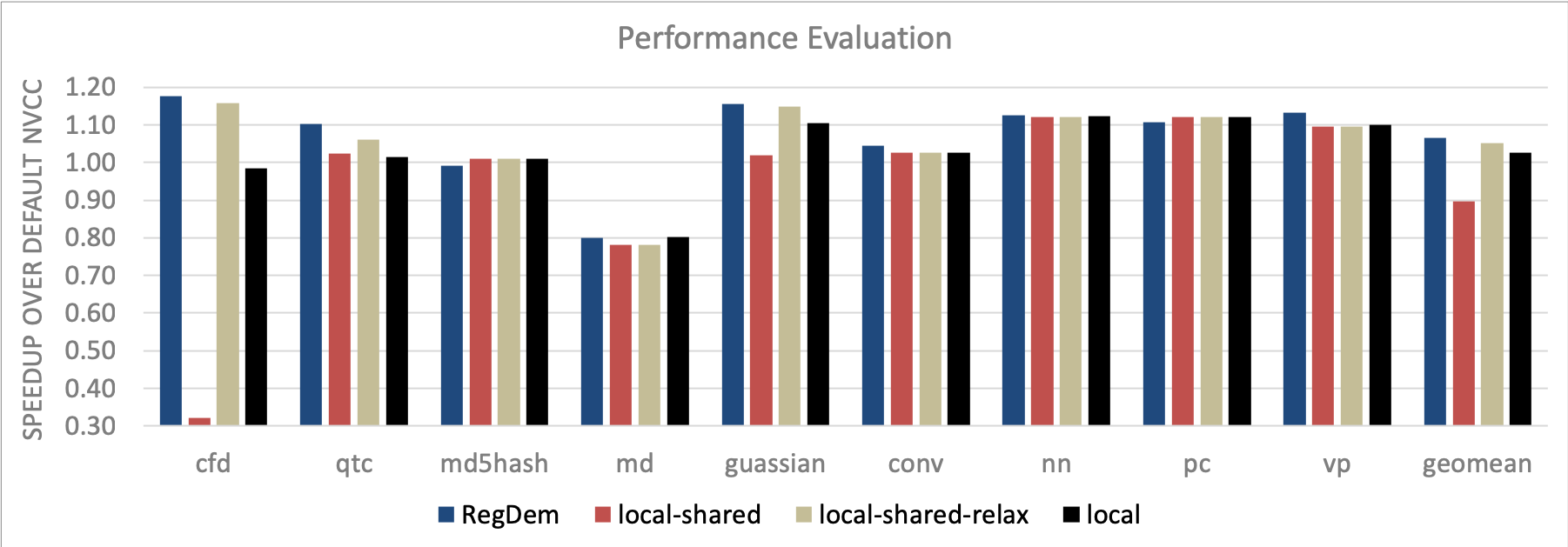}
\caption{Performance Evaluation of RegDem: The bars show speedups over default baseline \lstinline{nvcc} versions. The rightmost bars show the geometric mean speedup of various versions. RegDem uses the techniques presented in this paper. \emph{local}, \emph{local-shared}, and \emph{local-shared-relax} represent the alternative techniques for GPUs, with \emph{local-shared} being the closest alternative. Overall, RegDem obtains a 7\% geometric mean speedup over default \lstinline{nvcc} and outperforms the other alternative techniques in 7 benchmarks.} 
\label{fig:speedup}
\end{figure*}

Figure~\ref{fig:speedup} shows the speedups of RegDem and alternatives over the baseline \lstinline{nvcc} variants.
This experiment applies the best combination of optimization options presented in Section~\ref{sec:optimization}, found through exhaustive search over all combinations. 

Overall, RegDem performs the best in seven of the nine benchmarks when compared to other spilling techniques. RegDem achieves up to 1.18x speedup over the baseline \lstinline{nvcc}, with a geometric mean of 1.07x, and shows significant improvement over the closest research alternative (\emph{local-shared}), where RegDem obtains 1.19x geometric mean speedup.
In contrast, \emph{local}, \emph{local-shared}, \emph{local-shared-relax} achieve  1.03x, 0.90x, and 1.05x geometric mean speedups over the baseline implementations, respectively. 

\subsection{Discussion}
\label{sec:disc}
We classify the benchmarks into three groups based on their characteristics. The first group, \lstinline{cfd} and \lstinline{qtc}, requires a significant number of registers to be spilled. Local memory spilling is unable to improve the performance due to high access latencies. 
RegDem also benefits from the maximized single-thread performance~\cite{nvcc} of \lstinline{nvcc}'s default register allocation and shows substantial improvement over other spilling alternatives, which rely on the aggressive allocation.

The second group includes the remaining benchmarks except \lstinline{md}, wherein a few registers are spilled to reach the target occupancy. Hence, the spilled register access overhead is less noticeable. In several benchmarks, instead of spilling registers, \lstinline{nvcc} performs allocation in a manner that avoids spilling, but degrades single-thread performance (Table~\ref{tab:benchmarks-detail}). This results in an occupancy gain without spilling overhead, which we call \emph{zero spilling}. 

The \lstinline{md5hash} benchmark exemplifies the effect of zero spilling in the alternative approaches, resulting in slightly better performance than RegDem. However, zero spilling also comes with a drawback, as it sacrifices single-thread performance for occupancy. Such example is shown in \lstinline{vp}, where the performance achieved from zero spilling is below that of RegDem. Profiling results indicate that the number of dynamic instructions increased significantly (about 3\% per warp), owing to the reduced single-thread performance. Assembly inspection also showed that the additional instructions have high stall count (13 cycles). 

\begin{figure}[tb!]
\centering
\includegraphics[width=3.1in]{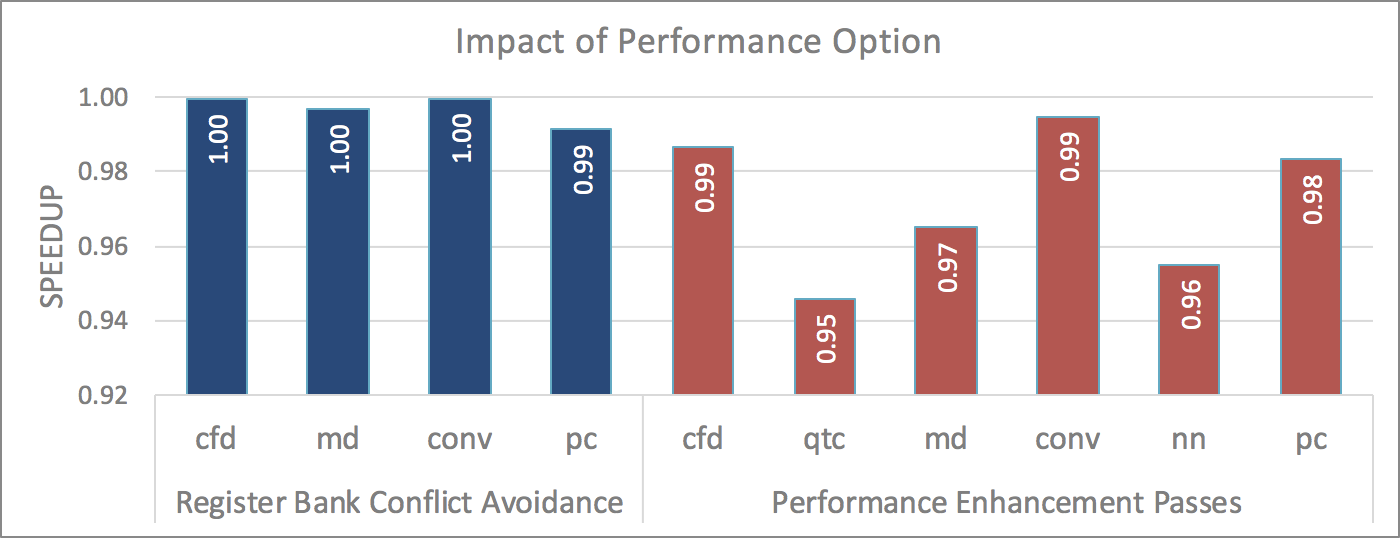}
\caption{Impact of the Post-Spilling Optimizations: Slowdown obtained by disabling individual performance options. A lower speedup indicates that the performance option has higher impact. On average, performance enhancement passes and register bank conflict avoidance show 3\% and less than 1\%  performance impact, respectively.} 
\label{fig:optimization-impact}
\end{figure}

The \lstinline{md} benchmark is the only benchmark that does not achieve improvement with any of the techniques. The key distinction of \lstinline{md} is that it uses double-precision floating point numbers, and hence the FP64 ALUs become the performance bottleneck. Improving occupancy by the described optimizations increases the execution time of the critical path, as more threads need to wait for the FP64 ALUs. 

\subsection{Impact of Post-Spilling Optimizations}
\label{sec:impact}
This section analyzes the impact of the performance options presented in Section~\ref{sec:optimization}. We used the best combination (\emph{RegDem} from Figure~\ref{fig:speedup}) as the baseline for the analysis. 

Figure~\ref{fig:optimization-impact} evaluates the register bank conflict avoidance and the performance enhancement passes. We measured the impact by disabling individual options and observing the performance change. Benchmarks that do not benefit from any option are not shown.   

Register bank conflict avoidance has an impact of less than 1\%. Although, \lstinline{MaxAs} reports that the algorithm could avoid an average of 36\% of the conflicts, the ratio of the instructions with conflict to the total number of instructions is low and therefore does not significantly affect program performance.

The performance enhancement pass improves the performance by up to 5\% and by approximately 3\% on average. We have observed that value register substitution is rarely employed. This stems from the rarity of code sections where free registers are available. Hence, only a small portion of the program can take advantage of the pass. 

\begin{figure}[tb!]
\centering
\includegraphics[width=3.1in]{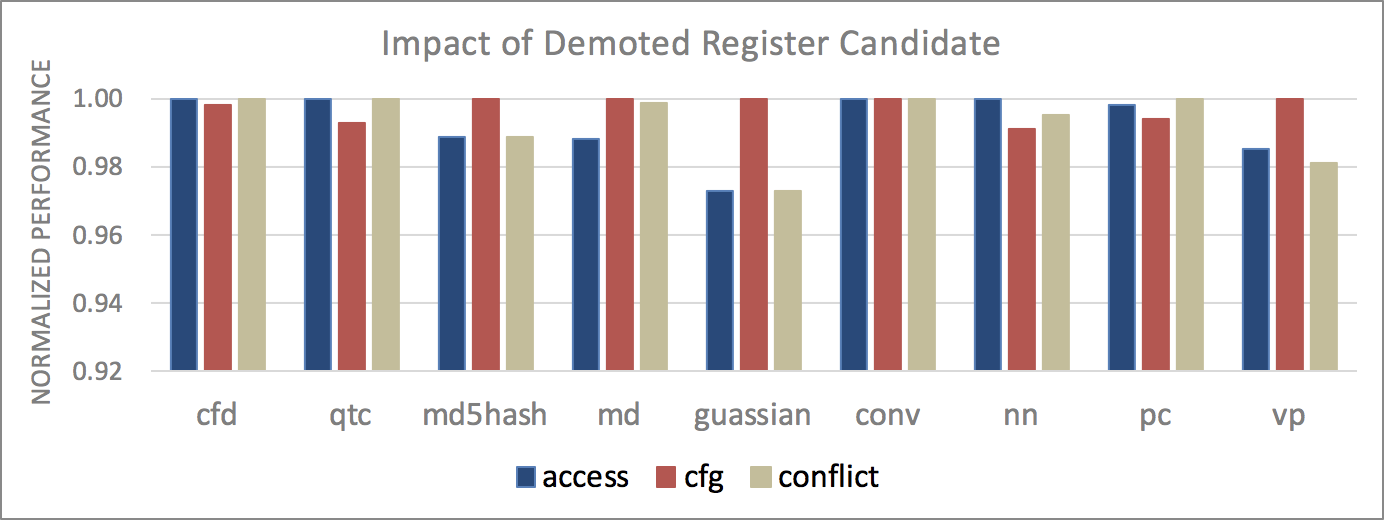}
\caption{Impact of Register Candidate: The chart shows normalized performance over the \emph{best} candidate selection strategy. The strategy with 1.0x speedup is the best strategy for that benchmark.Overall, the \emph{cfg} strategy obtains the best performance.} 
\label{fig:reg-candidate}
\end{figure}

We also evaluated the impact of choosing candidate registers for demotion, described in Section~\ref{sec:choosing-reg} (Figure~\ref{fig:reg-candidate}).  
Overall, the \emph{cfg} strategy gives the best result. The advantage of the \emph{cfg} approach is that it considers access overhead inside loops; however, it could result in over-estimating the overhead, and thus avoiding good candidates. 

\subsection{Compile-time Performance Prediction}
\label{sec:evalpredict}
This section evaluates the performance predictor. We compare the performance achieved by the predictor to an oracle that knows the best variant of the benchmark, including original, \emph{local}, \emph{local-shared}, and \emph{local-shared-relax} from Figure~\ref{fig:speedup}. 

The predictor estimates the performance of every code variant. For RegDem, the predictor estimates all performance option combinations, including the post-RegDem optimizations presented in Section~\ref{sec:optimization}. To break ties, it chooses the one with the highest number of performance options enabled, counting on potential benefits of the enabled options. We also compared the predictor with a naive scheme (\emph{naive}) that statically counts stall cycles, and RegDem with all performance enhancements enabled (\emph{RegDem}).

\begin{figure}[t!]
\centering
\includegraphics[width=3.1in]{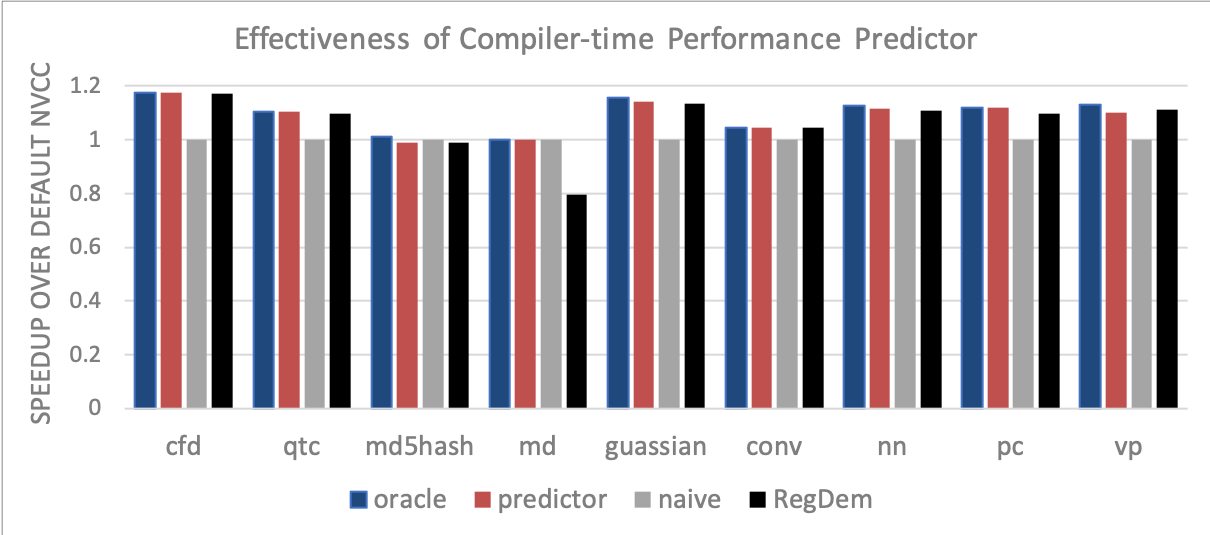}
\caption{Effectiveness of Compile-time Performance Predictor: The chart shows the speedup of each benchmark using an oracle and the performance predictor. The oracle and the predictor achieves a geometric mean speedup of 1.10x and 1.09x, respectively. Thus, the predictor can achieve 99.0\% performance compared to the oracle. }
\label{fig:estimation}
\end{figure}

Figure~\ref{fig:estimation} shows the results. The geometric mean speedup achieved by the oracle is 1.10x while the predictor obtains a geometric mean speedup of 1.09x over the default \lstinline{nvcc} versions, achieving 99.0\% of the results of exhaustive search. The predictor also helps avoid the worst-case scenario, where applying the optimization degrades the performance. 

In seven of the nine benchmarks, the predictor correctly chose the spilling technique with the best performance. We consider this good accuracy for static performance prediction. One notable case is the \lstinline{md} benchmark, which all optimizations fail to improve. The predictor correctly assesses the situation and chooses the low-occupancy variant.

The current predictor is biased toward optimizations that directly impact instruction stalls, such as removing and rescheduling memory instructions. This bias is due to limited consideration of instruction scheduling. A runtime method could improve the accuracy further, forfeiting the benefit of a compile-time solution. In future work, we will explore using microbenchmarks to build a database of instruction interactions, capturing effects of scheduling more accurately.

\section{Related Work}
\label{sec:related-work}

Generally, register spilling is performed during register allocation. Prior work has extensively studied strategies for CPUs~\cite{
graph-coloring,linear-scan, 
ssa-allocation,ssa-allocation-2,puzzle-allocation,graph-coloring-optimal,smith-register-allocation}. 
These algorithms aim for high register usage to maximize single-thread performance. Using the same approach on GPUs could lead to lower occupancy and reduced performance. 

Several approaches have been proposed to improve GPU register usage. Many of them exploit smarter register allocation~\cite{hayes-local-spill,hayes-orion,sampaio-reg-spill,xie-register-allocation,you-vector-register-alloc}. Similar to our work, Hayes and Zhang~\cite{hayes-local-spill} proposed a register allocation algorithm for GPUs that utilizes the shared memory for spilling. Their approach turns local memory spills into shared memory allocations. As shown in the evaluation section, this approach experiences reduced single-thread performance. Moreover, the approach does not handle multi-word data. Sampaio et al.~\cite{sampaio-reg-spill} proposed divergence-aware register allocation, which reduces register pressure by placing common data in non-register memory spaces. Xie et al.~\cite{xie-register-allocation} proposed a compiler framework that puts spilled registers into shared memory; however, their method applies the optimization at the PTX level and requires additional hardware support for register allocation. Hayes et al. further extended their work in Orion~\cite{hayes-orion} with a
register allocation scheme that spills registers to shared and local memory, instead of converting local memory spills. In contrast, RegDem does not fully reallocate the entire register space, which could interfere with transformations performed by prior optimization passes. Also their approach applies to the older Kepler ISA alone, and hence does not take instruction barriers into consideration. Additionally, our approach can be applied as a stand-alone optimization. 

Resource virtualization is another direction for reducing register pressure. Yan and Zhang~\cite{yan1,yan2} use virtual register files to realize such effect on CPUs. On GPUs, Zorau~\cite{zorau} uses resource virtualization to manage multiple on-chip resources, including registers and shared memory. Jeon et al.~\cite{jeon-register-file-virtualization} proposed register file virtualization to share physical registers across GPU warps but did not use shared memory as a spill target. These approaches require hardware support, which is unavailable in current GPUs. 

Prior work has proposed several auto-tuning systems for GPUs~\cite{openmpc,sabne12,david-auto-tuning,yuri-auto-tuning,sato-auto-tuning,script-auto-tuning,gadapt,jia-tuning}. However, the majority of these contributions rely on runtime information for performance tuning. Several authors have also studied offline performance analysis for GPUs. Baghsorkhi et al.~\cite{baghsorkhi-performance-analysis} proposed a work-flow graph for compiler-based performance analysis. Meng et al.~\cite{meng-performance-analysis} predict GPU performance from the CPU skeleton code. Unlike these two schemes, our performance predictor uses low-level information to estimate the performance of GPU programs. Sim et al.~\cite{sim-performance-analysis} proposed an offline performance analysis, which estimates the potential benefit to GPU programs. In contrast to our approach, Sim's method aims to find the bottleneck of the GPU program, while our scheme compares the performance of different code variants.  Our performance estimation can be seen as complementary to these approaches, allowing the analysis from multiple angles for higher-accuracy performance models. 

\section{Conclusion}
\label{sec:conclusion}

This paper proposed {\em RegDem:} an assembly-level GPU register optimization method for improving program occupancy by reducing register pressure. The optimization moves excessive registers to the on-chip shared memory, finding a good tradeoff between register use and occupancy. The optimization addresses issues such as bank conflicts and interactions with instruction scheduling. The paper also introduced three optimizations to further improve the resulting code of RegDem. The presented techniques work well in an automatic, stand-alone binary translator. Nevertheless, tighter integration with the \lstinline{nvcc} compiler could yield further improvements, especially through better interactions of register allocation, instruction scheduling, and instruction selection. Further opportunities lie in improving performance prediction through runtime methods. Such methods would need to carefully consider the overheads of making the decisions -- in this case selecting from among several code variants -- in the critical program execution path. The presented static method avoids these overheads and performs well in practice.

\bibliographystyle{ACM-Reference-Format}
\bibliography{references}


\begin{thebibliography}{00}


\ifx \showCODEN    \undefined \def \showCODEN     #1{\unskip}     \fi
\ifx \showDOI      \undefined \def \showDOI       #1{#1}\fi
\ifx \showISBNx    \undefined \def \showISBNx     #1{\unskip}     \fi
\ifx \showISBNxiii \undefined \def \showISBNxiii  #1{\unskip}     \fi
\ifx \showISSN     \undefined \def \showISSN      #1{\unskip}     \fi
\ifx \showLCCN     \undefined \def \showLCCN      #1{\unskip}     \fi
\ifx \shownote     \undefined \def \shownote      #1{#1}          \fi
\ifx \showarticletitle \undefined \def \showarticletitle #1{#1}   \fi
\ifx \showURL      \undefined \def \showURL       {\relax}        \fi
\providecommand\bibfield[2]{#2}
\providecommand\bibinfo[2]{#2}
\providecommand\natexlab[1]{#1}
\providecommand\showeprint[2][]{arXiv:#2}

\bibitem[\protect\citeauthoryear{Baghsorkhi, Delahaye, Patel, Gropp, and
  Hwu}{Baghsorkhi et~al\mbox{.}}{2010}]%
        {baghsorkhi-performance-analysis}
\bibfield{author}{\bibinfo{person}{Sara~S. Baghsorkhi},
  \bibinfo{person}{Matthieu Delahaye}, \bibinfo{person}{Sanjay~J. Patel},
  \bibinfo{person}{William~D. Gropp}, {and} \bibinfo{person}{Wen-mei~W. Hwu}.}
  \bibinfo{year}{2010}\natexlab{}.
\newblock \showarticletitle{An Adaptive Performance Modeling Tool for GPU
  Architectures}.
\newblock \bibinfo{journal}{{\em SIGPLAN Not.\/}} \bibinfo{volume}{45},
  \bibinfo{number}{5} (\bibinfo{date}{Jan.} \bibinfo{year}{2010}),
  \bibinfo{pages}{105--114}.
\newblock
\showISSN{0362-1340}
\showDOI{%
\url{https://doi.org/10.1145/1837853.1693470}}


\bibitem[\protect\citeauthoryear{Briggs, Cooper, and Torczon}{Briggs
  et~al\mbox{.}}{1992}]%
        {Briggs:1992}
\bibfield{author}{\bibinfo{person}{Preston Briggs}, \bibinfo{person}{Keith~D.
  Cooper}, {and} \bibinfo{person}{Linda Torczon}.}
  \bibinfo{year}{1992}\natexlab{}.
\newblock \showarticletitle{Rematerialization}.
\newblock \bibinfo{journal}{{\em SIGPLAN Not.\/}} \bibinfo{volume}{27},
  \bibinfo{number}{7} (\bibinfo{date}{July} \bibinfo{year}{1992}),
  \bibinfo{pages}{311--321}.
\newblock
\showISSN{0362-1340}
\showDOI{%
\url{https://doi.org/10.1145/143103.143143}}


\bibitem[\protect\citeauthoryear{Chaitin}{Chaitin}{1982}]%
        {graph-coloring}
\bibfield{author}{\bibinfo{person}{G.~J. Chaitin}.}
  \bibinfo{year}{1982}\natexlab{}.
\newblock \showarticletitle{Register Allocation \& Spilling via Graph
  Coloring}. In \bibinfo{booktitle}{{\em Proceedings of the 1982 SIGPLAN
  Symposium on Compiler Construction}} {\em (\bibinfo{series}{SIGPLAN '82})}.
  \bibinfo{publisher}{ACM}, \bibinfo{address}{New York, NY, USA},
  \bibinfo{pages}{98--105}.
\newblock
\showISBNx{0-89791-074-5}
\showDOI{%
\url{https://doi.org/10.1145/800230.806984}}


\bibitem[\protect\citeauthoryear{Che, Sheaffer, Boyer, Szafaryn, Wang, and
  Skadron}{Che et~al\mbox{.}}{2010}]%
        {rodinia-benchmark}
\bibfield{author}{\bibinfo{person}{Shuai Che}, \bibinfo{person}{Jeremy~W.
  Sheaffer}, \bibinfo{person}{Michael Boyer}, \bibinfo{person}{Lukasz~G.
  Szafaryn}, \bibinfo{person}{Liang Wang}, {and} \bibinfo{person}{Kevin
  Skadron}.} \bibinfo{year}{2010}\natexlab{}.
\newblock \showarticletitle{A Characterization of the Rodinia Benchmark Suite
  with Comparison to Contemporary CMP Workloads}. In \bibinfo{booktitle}{{\em
  Proceedings of the IEEE International Symposium on Workload Characterization
  (IISWC'10)}} {\em (\bibinfo{series}{IISWC '10})}. \bibinfo{publisher}{IEEE
  Computer Society}, \bibinfo{address}{Washington, DC, USA},
  \bibinfo{pages}{1--11}.
\newblock
\showISBNx{978-1-4244-9297-8}
\showDOI{%
\url{https://doi.org/10.1109/IISWC.2010.5650274}}


\bibitem[\protect\citeauthoryear{Danalis, Marin, McCurdy, Meredith, Roth,
  Spafford, Tipparaju, and Vetter}{Danalis et~al\mbox{.}}{2010}]%
        {shoc-benchmark}
\bibfield{author}{\bibinfo{person}{Anthony Danalis}, \bibinfo{person}{Gabriel
  Marin}, \bibinfo{person}{Collin McCurdy}, \bibinfo{person}{Jeremy~S.
  Meredith}, \bibinfo{person}{Philip~C. Roth}, \bibinfo{person}{Kyle Spafford},
  \bibinfo{person}{Vinod Tipparaju}, {and} \bibinfo{person}{Jeffrey~S.
  Vetter}.} \bibinfo{year}{2010}\natexlab{}.
\newblock \showarticletitle{The Scalable Heterogeneous Computing (SHOC)
  Benchmark Suite}. In \bibinfo{booktitle}{{\em Proceedings of the 3rd Workshop
  on General-Purpose Computation on Graphics Processing Units}} {\em
  (\bibinfo{series}{GPGPU-3})}. \bibinfo{publisher}{ACM}, \bibinfo{address}{New
  York, NY, USA}, \bibinfo{pages}{63--74}.
\newblock
\showISBNx{978-1-60558-935-0}
\showDOI{%
\url{https://doi.org/10.1145/1735688.1735702}}


\bibitem[\protect\citeauthoryear{Davidson and Owens}{Davidson and
  Owens}{2012}]%
        {david-auto-tuning}
\bibfield{author}{\bibinfo{person}{Andrew Davidson} {and} \bibinfo{person}{John
  Owens}.} \bibinfo{year}{2012}\natexlab{}.
\newblock \bibinfo{booktitle}{{\em Toward Techniques for Auto-tuning GPU
  Algorithms}}.
\newblock \bibinfo{publisher}{Springer Berlin Heidelberg},
  \bibinfo{address}{Berlin, Heidelberg}, \bibinfo{pages}{110--119}.
\newblock
\showISBNx{978-3-642-28145-7}
\showDOI{%
\url{https://doi.org/10.1007/978-3-642-28145-7_11}}


\bibitem[\protect\citeauthoryear{Dotsenko, Baghsorkhi, Lloyd, and
  Govindaraju}{Dotsenko et~al\mbox{.}}{2011}]%
        {yuri-auto-tuning}
\bibfield{author}{\bibinfo{person}{Yuri Dotsenko}, \bibinfo{person}{Sara~S.
  Baghsorkhi}, \bibinfo{person}{Brandon Lloyd}, {and} \bibinfo{person}{Naga~K.
  Govindaraju}.} \bibinfo{year}{2011}\natexlab{}.
\newblock \showarticletitle{Auto-tuning of Fast Fourier Transform on Graphics
  Processors}.
\newblock \bibinfo{journal}{{\em SIGPLAN Not.\/}} \bibinfo{volume}{46},
  \bibinfo{number}{8} (\bibinfo{date}{Feb.} \bibinfo{year}{2011}),
  \bibinfo{pages}{257--266}.
\newblock
\showISSN{0362-1340}
\showDOI{%
\url{https://doi.org/10.1145/2038037.1941589}}


\bibitem[\protect\citeauthoryear{Gray}{Gray}{2017}]%
        {maxas}
\bibfield{author}{\bibinfo{person}{Scott Gray}.}
  \bibinfo{year}{2017}\natexlab{}.
\newblock \bibinfo{title}{{MaxAs}}.
\newblock
  \bibinfo{howpublished}{\url{https://github.com/NervanaSystems/maxas/}}.
  (\bibinfo{year}{2017}).
\newblock
\newblock
\shownote{[Online; accessed 1-April-2017].}


\bibitem[\protect\citeauthoryear{Hack, Grund, and Goos}{Hack
  et~al\mbox{.}}{2006}]%
        {ssa-allocation}
\bibfield{author}{\bibinfo{person}{Sebastian Hack}, \bibinfo{person}{Daniel
  Grund}, {and} \bibinfo{person}{Gerhard Goos}.}
  \bibinfo{year}{2006}\natexlab{}.
\newblock \showarticletitle{Register Allocation for Programs in SSA-Form}. In
  \bibinfo{booktitle}{{\em Proceedings of the 15th International Conference on
  Compiler Construction}} {\em (\bibinfo{series}{CC'06})}.
  \bibinfo{publisher}{Springer-Verlag}, \bibinfo{address}{Berlin, Heidelberg},
  \bibinfo{pages}{247--262}.
\newblock
\showISBNx{3-540-33050-X, 978-3-540-33050-9}
\showDOI{%
\url{https://doi.org/10.1007/11688839_20}}


\bibitem[\protect\citeauthoryear{Hayes, Li, Chavarr\'{\i}a-Miranda, Song, and
  Zhang}{Hayes et~al\mbox{.}}{2016}]%
        {hayes-orion}
\bibfield{author}{\bibinfo{person}{Ari~B. Hayes}, \bibinfo{person}{Lingda Li},
  \bibinfo{person}{Daniel Chavarr\'{\i}a-Miranda},
  \bibinfo{person}{Shuaiwen~Leon Song}, {and} \bibinfo{person}{Eddy~Z. Zhang}.}
  \bibinfo{year}{2016}\natexlab{}.
\newblock \showarticletitle{Orion: A Framework for GPU Occupancy Tuning}. In
  \bibinfo{booktitle}{{\em Proceedings of the 17th International Middleware
  Conference}} {\em (\bibinfo{series}{Middleware '16})}.
  \bibinfo{publisher}{ACM}, \bibinfo{address}{New York, NY, USA}, Article
  \bibinfo{articleno}{18}, \bibinfo{numpages}{13}~pages.
\newblock
\showISBNx{978-1-4503-4300-8}
\showDOI{%
\url{https://doi.org/10.1145/2988336.2988355}}


\bibitem[\protect\citeauthoryear{Hayes and Zhang}{Hayes and Zhang}{2014}]%
        {hayes-local-spill}
\bibfield{author}{\bibinfo{person}{Ari~B. Hayes} {and} \bibinfo{person}{Eddy~Z.
  Zhang}.} \bibinfo{year}{2014}\natexlab{}.
\newblock \showarticletitle{Unified On-chip Memory Allocation for SIMT
  Architecture}. In \bibinfo{booktitle}{{\em Proceedings of the 28th ACM
  International Conference on Supercomputing}} {\em (\bibinfo{series}{ICS
  '14})}. \bibinfo{publisher}{ACM}, \bibinfo{address}{New York, NY, USA},
  \bibinfo{pages}{293--302}.
\newblock
\showISBNx{978-1-4503-2642-1}
\showDOI{%
\url{https://doi.org/10.1145/2597652.2597685}}


\bibitem[\protect\citeauthoryear{Jeon, Ravi, Kim, and Annavaram}{Jeon
  et~al\mbox{.}}{2015}]%
        {jeon-register-file-virtualization}
\bibfield{author}{\bibinfo{person}{Hyeran Jeon},
  \bibinfo{person}{Gokul~Subramanian Ravi}, \bibinfo{person}{Nam~Sung Kim},
  {and} \bibinfo{person}{Murali Annavaram}.} \bibinfo{year}{2015}\natexlab{}.
\newblock \showarticletitle{GPU Register File Virtualization}. In
  \bibinfo{booktitle}{{\em Proceedings of the 48th International Symposium on
  Microarchitecture}} {\em (\bibinfo{series}{MICRO-48})}.
  \bibinfo{publisher}{ACM}, \bibinfo{address}{New York, NY, USA},
  \bibinfo{pages}{420--432}.
\newblock
\showISBNx{978-1-4503-4034-2}
\showDOI{%
\url{https://doi.org/10.1145/2830772.2830784}}


\bibitem[\protect\citeauthoryear{Jia, Garza, Shaw, and Martonosi}{Jia
  et~al\mbox{.}}{2015}]%
        {jia-tuning}
\bibfield{author}{\bibinfo{person}{Wenhao Jia}, \bibinfo{person}{Elba Garza},
  \bibinfo{person}{Kelly~A. Shaw}, {and} \bibinfo{person}{Margaret Martonosi}.}
  \bibinfo{year}{2015}\natexlab{}.
\newblock \showarticletitle{GPU Performance and Power Tuning Using Regression
  Trees}.
\newblock \bibinfo{journal}{{\em ACM Trans. Archit. Code Optim.\/}}
  \bibinfo{volume}{12}, \bibinfo{number}{2}, Article \bibinfo{articleno}{13}
  (\bibinfo{date}{May} \bibinfo{year}{2015}), \bibinfo{numpages}{26}~pages.
\newblock
\showISSN{1544-3566}
\showDOI{%
\url{https://doi.org/10.1145/2736287}}


\bibitem[\protect\citeauthoryear{Jia, Maggioni, Staiger, and Scarpazza}{Jia
  et~al\mbox{.}}{2018}]%
        {volta-architecture}
\bibfield{author}{\bibinfo{person}{Zhe Jia}, \bibinfo{person}{Marco Maggioni},
  \bibinfo{person}{Benjamin Staiger}, {and} \bibinfo{person}{Daniele~Paolo
  Scarpazza}.} \bibinfo{year}{2018}\natexlab{}.
\newblock \showarticletitle{Dissecting the {NVIDIA} Volta {GPU} Architecture
  via Microbenchmarking}.
\newblock \bibinfo{journal}{{\em CoRR\/}}  \bibinfo{volume}{abs/1804.06826}
  (\bibinfo{year}{2018}).
\newblock
\showeprint[arxiv]{1804.06826}
\showURL{%
\url{http://arxiv.org/abs/1804.06826}}


\bibitem[\protect\citeauthoryear{Khan, Basu, Rudy, Hall, Chen, and Chame}{Khan
  et~al\mbox{.}}{2013}]%
        {script-auto-tuning}
\bibfield{author}{\bibinfo{person}{Malik Khan}, \bibinfo{person}{Protonu Basu},
  \bibinfo{person}{Gabe Rudy}, \bibinfo{person}{Mary Hall},
  \bibinfo{person}{Chun Chen}, {and} \bibinfo{person}{Jacqueline Chame}.}
  \bibinfo{year}{2013}\natexlab{}.
\newblock \showarticletitle{A Script-based Autotuning Compiler System to
  Generate High-performance CUDA Code}.
\newblock \bibinfo{journal}{{\em ACM Trans. Archit. Code Optim.\/}}
  \bibinfo{volume}{9}, \bibinfo{number}{4}, Article \bibinfo{articleno}{31}
  (\bibinfo{date}{Jan.} \bibinfo{year}{2013}), \bibinfo{numpages}{25}~pages.
\newblock
\showISSN{1544-3566}
\showDOI{%
\url{https://doi.org/10.1145/2400682.2400690}}


\bibitem[\protect\citeauthoryear{Krause}{Krause}{2013}]%
        {graph-coloring-optimal}
\bibfield{author}{\bibinfo{person}{Philipp~Klaus Krause}.}
  \bibinfo{year}{2013}\natexlab{}.
\newblock \bibinfo{booktitle}{{\em Optimal Register Allocation in Polynomial
  Time}}.
\newblock \bibinfo{publisher}{Springer Berlin Heidelberg},
  \bibinfo{address}{Berlin, Heidelberg}, \bibinfo{pages}{1--20}.
\newblock
\showISBNx{978-3-642-37051-9}
\showDOI{%
\url{https://doi.org/10.1007/978-3-642-37051-9_1}}


\bibitem[\protect\citeauthoryear{Lee and Eigenmann}{Lee and Eigenmann}{2010}]%
        {openmpc}
\bibfield{author}{\bibinfo{person}{S. Lee} {and} \bibinfo{person}{R.
  Eigenmann}.} \bibinfo{year}{2010}\natexlab{}.
\newblock \showarticletitle{OpenMPC: Extended OpenMP Programming and Tuning for
  GPUs}. In \bibinfo{booktitle}{{\em 2010 ACM/IEEE International Conference for
  High Performance Computing, Networking, Storage and Analysis}}.
  \bibinfo{pages}{1--11}.
\newblock
\showISSN{2167-4329}
\showDOI{%
\url{https://doi.org/10.1109/SC.2010.36}}


\bibitem[\protect\citeauthoryear{Liu, Hegde, and Kulkarni}{Liu
  et~al\mbox{.}}{2016}]%
        {fsm-benchmark}
\bibfield{author}{\bibinfo{person}{Jianqiao Liu}, \bibinfo{person}{Nikhil
  Hegde}, {and} \bibinfo{person}{Milind Kulkarni}.}
  \bibinfo{year}{2016}\natexlab{}.
\newblock \showarticletitle{Hybrid {CPU-GPU} scheduling and execution of tree
  traversals}. In \bibinfo{booktitle}{{\em Proceedings of the 2016
  International Conference on Supercomputing, {ICS} 2016, Istanbul, Turkey,
  June 1-3, 2016}}. \bibinfo{pages}{2:1--2:12}.
\newblock
\showDOI{%
\url{https://doi.org/10.1145/2925426.2926261}}


\bibitem[\protect\citeauthoryear{Liu, Zhang, and Shen}{Liu
  et~al\mbox{.}}{2009}]%
        {gadapt}
\bibfield{author}{\bibinfo{person}{Yixun Liu}, \bibinfo{person}{E.~Z. Zhang},
  {and} \bibinfo{person}{X. Shen}.} \bibinfo{year}{2009}\natexlab{}.
\newblock \showarticletitle{A cross-input adaptive framework for GPU program
  optimizations}. In \bibinfo{booktitle}{{\em 2009 IEEE International Symposium
  on Parallel Distributed Processing}}. \bibinfo{pages}{1--10}.
\newblock
\showISSN{1530-2075}
\showDOI{%
\url{https://doi.org/10.1109/IPDPS.2009.5160988}}


\bibitem[\protect\citeauthoryear{Meng, Morozov, Kumaran, Vishwanath, and
  Uram}{Meng et~al\mbox{.}}{2011}]%
        {meng-performance-analysis}
\bibfield{author}{\bibinfo{person}{J. Meng}, \bibinfo{person}{V.~A. Morozov},
  \bibinfo{person}{K. Kumaran}, \bibinfo{person}{V. Vishwanath}, {and}
  \bibinfo{person}{T.~D. Uram}.} \bibinfo{year}{2011}\natexlab{}.
\newblock \showarticletitle{GROPHECY: GPU performance projection from CPU code
  skeletons}. In \bibinfo{booktitle}{{\em 2011 International Conference for
  High Performance Computing, Networking, Storage and Analysis (SC)}}.
  \bibinfo{pages}{1--11}.
\newblock
\showISSN{2167-4329}


\bibitem[\protect\citeauthoryear{{NVIDIA}}{{NVIDIA}}{2017a}]%
        {cuda-best-practice}
\bibfield{author}{\bibinfo{person}{{NVIDIA}}.}
  \bibinfo{year}{2017}\natexlab{a}.
\newblock \bibinfo{title}{{CUDA C} Best Practices Guide}.
\newblock
  \bibinfo{howpublished}{\url{http://docs.nvidia.com/cuda/cuda-c-best-practices-guide}}.
    (\bibinfo{year}{2017}).
\newblock
\newblock
\shownote{[Online; accessed 2-April-2017].}


\bibitem[\protect\citeauthoryear{{NVIDIA}}{{NVIDIA}}{2017b}]%
        {cuda-programming-guide}
\bibfield{author}{\bibinfo{person}{{NVIDIA}}.}
  \bibinfo{year}{2017}\natexlab{b}.
\newblock \bibinfo{title}{{CUDA C} Programming Guide}.
\newblock
  \bibinfo{howpublished}{\url{http://docs.nvidia.com/cuda/cuda-c-programming-guide/}}.
    (\bibinfo{year}{2017}).
\newblock
\newblock
\shownote{[Online; accessed 2-April-2017].}


\bibitem[\protect\citeauthoryear{{NVIDIA}}{{NVIDIA}}{2017c}]%
        {cuda-occupancy-calc}
\bibfield{author}{\bibinfo{person}{{NVIDIA}}.}
  \bibinfo{year}{2017}\natexlab{c}.
\newblock \bibinfo{title}{{CUDA} Occupancy Calculator}.
\newblock
  \bibinfo{howpublished}{\url{https://developer.download.nvidia.com/compute/cuda/CUDA_Occupancy_calculator.xls}}.
    (\bibinfo{year}{2017}).
\newblock
\newblock
\shownote{[Online; accessed 9-April-2018].}


\bibitem[\protect\citeauthoryear{{NVIDIA}}{{NVIDIA}}{2017d}]%
        {cuda-toolkit}
\bibfield{author}{\bibinfo{person}{{NVIDIA}}.}
  \bibinfo{year}{2017}\natexlab{d}.
\newblock \bibinfo{title}{{CUDA} Toolkit Documentation - {CUDA} Samples}.
\newblock
  \bibinfo{howpublished}{\url{http://docs.nvidia.com/cuda/cuda-samples}}.
  (\bibinfo{year}{2017}).
\newblock
\newblock
\shownote{[Online; accessed 1-April-2017].}


\bibitem[\protect\citeauthoryear{{NVIDIA}}{{NVIDIA}}{2017e}]%
        {nvcc}
\bibfield{author}{\bibinfo{person}{{NVIDIA}}.}
  \bibinfo{year}{2017}\natexlab{e}.
\newblock \bibinfo{title}{{NVIDIA} {CUDA} Compiler Driver {NVCC}}.
\newblock
  \bibinfo{howpublished}{\url{http://docs.nvidia.com/cuda/cuda-compiler-driver-nvcc/}}.
    (\bibinfo{year}{2017}).
\newblock
\newblock
\shownote{[Online; accessed 2-April-2017].}


\bibitem[\protect\citeauthoryear{Podlozhnyuk}{Podlozhnyuk}{2013}]%
        {cuda-conv}
\bibfield{author}{\bibinfo{person}{Victor Podlozhnyuk}.}
  \bibinfo{year}{2013}\natexlab{}.
\newblock \bibinfo{booktitle}{{\em Image Convolution with CUDA}}.
\newblock \bibinfo{type}{{T}echnical {R}eport}. \bibinfo{institution}{NVIDIA
  Corporation}.
\newblock


\bibitem[\protect\citeauthoryear{Poletto and Sarkar}{Poletto and
  Sarkar}{1999}]%
        {linear-scan}
\bibfield{author}{\bibinfo{person}{Massimiliano Poletto} {and}
  \bibinfo{person}{Vivek Sarkar}.} \bibinfo{year}{1999}\natexlab{}.
\newblock \showarticletitle{Linear Scan Register Allocation}.
\newblock \bibinfo{journal}{{\em ACM Trans. Program. Lang. Syst.\/}}
  \bibinfo{volume}{21}, \bibinfo{number}{5} (\bibinfo{date}{Sept.}
  \bibinfo{year}{1999}), \bibinfo{pages}{895--913}.
\newblock
\showISSN{0164-0925}
\showDOI{%
\url{https://doi.org/10.1145/330249.330250}}


\bibitem[\protect\citeauthoryear{Quint\~{a}o Pereira and Palsberg}{Quint\~{a}o
  Pereira and Palsberg}{2008}]%
        {puzzle-allocation}
\bibfield{author}{\bibinfo{person}{Fernando~Magno Quint\~{a}o Pereira} {and}
  \bibinfo{person}{Jens Palsberg}.} \bibinfo{year}{2008}\natexlab{}.
\newblock \showarticletitle{Register Allocation by Puzzle Solving}.
\newblock \bibinfo{journal}{{\em SIGPLAN Not.\/}} \bibinfo{volume}{43},
  \bibinfo{number}{6} (\bibinfo{date}{June} \bibinfo{year}{2008}),
  \bibinfo{pages}{216--226}.
\newblock
\showISSN{0362-1340}
\showDOI{%
\url{https://doi.org/10.1145/1379022.1375609}}


\bibitem[\protect\citeauthoryear{Sabne, Sakdhnagool, and Eigenmann}{Sabne
  et~al\mbox{.}}{2012}]%
        {sabne12}
\bibfield{author}{\bibinfo{person}{Amit Sabne}, \bibinfo{person}{Putt
  Sakdhnagool}, {and} \bibinfo{person}{Rudolf Eigenmann}.}
  \bibinfo{year}{2012}\natexlab{}.
\newblock \showarticletitle{Effects of Compiler Optimizations in OpenMP to CUDA
  Translation}. In \bibinfo{booktitle}{{\em Proc. of the International Workshop
  on OpenMP, IWOMP}}.
\newblock
\showURL{%
\url{http://engineering.purdue.edu/paramnt/publications/iwomp12.pdf}}


\bibitem[\protect\citeauthoryear{Sampaio, Gedeon, Pereira, and
  Collange}{Sampaio et~al\mbox{.}}{2012}]%
        {sampaio-reg-spill}
\bibfield{author}{\bibinfo{person}{Diogo~Nunes Sampaio}, \bibinfo{person}{Elie
  Gedeon}, \bibinfo{person}{Fernando Magno~Quint{\~a}o Pereira}, {and}
  \bibinfo{person}{Sylvain Collange}.} \bibinfo{year}{2012}\natexlab{}.
\newblock \bibinfo{booktitle}{{\em Spill Code Placement for SIMD Machines}}.
\newblock \bibinfo{publisher}{Springer Berlin Heidelberg},
  \bibinfo{address}{Berlin, Heidelberg}, \bibinfo{pages}{12--26}.
\newblock
\showISBNx{978-3-642-33182-4}
\showDOI{%
\url{https://doi.org/10.1007/978-3-642-33182-4_3}}


\bibitem[\protect\citeauthoryear{Sato, Takizawa, Komatsu, and Kobayashi}{Sato
  et~al\mbox{.}}{2010}]%
        {sato-auto-tuning}
\bibfield{author}{\bibinfo{person}{Katsuto Sato}, \bibinfo{person}{Hiroyuki
  Takizawa}, \bibinfo{person}{Kazuhiko Komatsu}, {and} \bibinfo{person}{Hiroaki
  Kobayashi}.} \bibinfo{year}{2010}\natexlab{}.
\newblock \bibinfo{booktitle}{{\em Automatic Tuning of CUDA Execution
  Parameters for Stencil Processing}}.
\newblock \bibinfo{publisher}{Springer New York}, \bibinfo{address}{New York,
  NY}, \bibinfo{pages}{209--228}.
\newblock
\showISBNx{978-1-4419-6935-4}
\showDOI{%
\url{https://doi.org/10.1007/978-1-4419-6935-4_13}}


\bibitem[\protect\citeauthoryear{Sim, Dasgupta, Kim, and Vuduc}{Sim
  et~al\mbox{.}}{2012}]%
        {sim-performance-analysis}
\bibfield{author}{\bibinfo{person}{Jaewoong Sim}, \bibinfo{person}{Aniruddha
  Dasgupta}, \bibinfo{person}{Hyesoon Kim}, {and} \bibinfo{person}{Richard
  Vuduc}.} \bibinfo{year}{2012}\natexlab{}.
\newblock \showarticletitle{A Performance Analysis Framework for Identifying
  Potential Benefits in GPGPU Applications}. In \bibinfo{booktitle}{{\em
  Proceedings of the 17th ACM SIGPLAN Symposium on Principles and Practice of
  Parallel Programming}} {\em (\bibinfo{series}{PPoPP '12})}.
  \bibinfo{publisher}{ACM}, \bibinfo{address}{New York, NY, USA},
  \bibinfo{pages}{11--22}.
\newblock
\showISBNx{978-1-4503-1160-1}
\showDOI{%
\url{https://doi.org/10.1145/2145816.2145819}}


\bibitem[\protect\citeauthoryear{Smith, Ramsey, and Holloway}{Smith
  et~al\mbox{.}}{2004}]%
        {smith-register-allocation}
\bibfield{author}{\bibinfo{person}{Michael~D. Smith}, \bibinfo{person}{Norman
  Ramsey}, {and} \bibinfo{person}{Glenn Holloway}.}
  \bibinfo{year}{2004}\natexlab{}.
\newblock \showarticletitle{A Generalized Algorithm for Graph-coloring Register
  Allocation}. In \bibinfo{booktitle}{{\em Proceedings of the ACM SIGPLAN 2004
  Conference on Programming Language Design and Implementation}} {\em
  (\bibinfo{series}{PLDI '04})}. \bibinfo{publisher}{ACM},
  \bibinfo{address}{New York, NY, USA}, \bibinfo{pages}{277--288}.
\newblock
\showISBNx{1-58113-807-5}
\showDOI{%
\url{https://doi.org/10.1145/996841.996875}}


\bibitem[\protect\citeauthoryear{Vijaykumar, Hsieh, Pekhimenko, Khan, Shrestha,
  Ghose, Jog, Gibbons, and Mutlu}{Vijaykumar et~al\mbox{.}}{2016}]%
        {zorau}
\bibfield{author}{\bibinfo{person}{N. Vijaykumar}, \bibinfo{person}{K. Hsieh},
  \bibinfo{person}{G. Pekhimenko}, \bibinfo{person}{S. Khan},
  \bibinfo{person}{A. Shrestha}, \bibinfo{person}{S. Ghose},
  \bibinfo{person}{A. Jog}, \bibinfo{person}{P.~B. Gibbons}, {and}
  \bibinfo{person}{O. Mutlu}.} \bibinfo{year}{2016}\natexlab{}.
\newblock \showarticletitle{Zorua: A holistic approach to resource
  virtualization in GPUs}. In \bibinfo{booktitle}{{\em 2016 49th Annual
  IEEE/ACM International Symposium on Microarchitecture (MICRO)}}.
  \bibinfo{pages}{1--14}.
\newblock
\showDOI{%
\url{https://doi.org/10.1109/MICRO.2016.7783718}}


\bibitem[\protect\citeauthoryear{Volkov}{Volkov}{2010}]%
        {volkov-occupancy}
\bibfield{author}{\bibinfo{person}{V. Volkov}.}
  \bibinfo{year}{2010}\natexlab{}.
\newblock \showarticletitle{Better performance at lower occupancy.}. In
  \bibinfo{booktitle}{{\em Proceedings of the GPU Technology Conference}}.
\newblock


\bibitem[\protect\citeauthoryear{Wimmer and Franz}{Wimmer and Franz}{2010}]%
        {ssa-allocation-2}
\bibfield{author}{\bibinfo{person}{Christian Wimmer} {and}
  \bibinfo{person}{Michael Franz}.} \bibinfo{year}{2010}\natexlab{}.
\newblock \showarticletitle{Linear Scan Register Allocation on SSA Form}. In
  \bibinfo{booktitle}{{\em Proceedings of the 8th Annual IEEE/ACM International
  Symposium on Code Generation and Optimization}} {\em (\bibinfo{series}{CGO
  '10})}. \bibinfo{publisher}{ACM}, \bibinfo{address}{New York, NY, USA},
  \bibinfo{pages}{170--179}.
\newblock
\showISBNx{978-1-60558-635-9}
\showDOI{%
\url{https://doi.org/10.1145/1772954.1772979}}


\bibitem[\protect\citeauthoryear{Xie, Liang, Li, Wu, Sun, Wang, and Fan}{Xie
  et~al\mbox{.}}{2015}]%
        {xie-register-allocation}
\bibfield{author}{\bibinfo{person}{Xiaolong Xie}, \bibinfo{person}{Yun Liang},
  \bibinfo{person}{Xiuhong Li}, \bibinfo{person}{Yudong Wu},
  \bibinfo{person}{Guangyu Sun}, \bibinfo{person}{Tao Wang}, {and}
  \bibinfo{person}{Dongrui Fan}.} \bibinfo{year}{2015}\natexlab{}.
\newblock \showarticletitle{Enabling Coordinated Register Allocation and
  Thread-level Parallelism Optimization for GPUs}. In \bibinfo{booktitle}{{\em
  Proceedings of the 48th International Symposium on Microarchitecture}} {\em
  (\bibinfo{series}{MICRO-48})}. \bibinfo{publisher}{ACM},
  \bibinfo{address}{New York, NY, USA}, \bibinfo{pages}{395--406}.
\newblock
\showISBNx{978-1-4503-4034-2}
\showDOI{%
\url{https://doi.org/10.1145/2830772.2830813}}


\bibitem[\protect\citeauthoryear{Yan and Zhang}{Yan and Zhang}{2007}]%
        {yan1}
\bibfield{author}{\bibinfo{person}{Jun Yan} {and} \bibinfo{person}{Wei Zhang}.}
  \bibinfo{year}{2007}\natexlab{}.
\newblock \bibinfo{booktitle}{{\em Virtual Registers: Reducing Register
  Pressure Without Enlarging the Register File}}.
\newblock \bibinfo{publisher}{Springer Berlin Heidelberg},
  \bibinfo{address}{Berlin, Heidelberg}, \bibinfo{pages}{57--70}.
\newblock
\showISBNx{978-3-540-69338-3}
\showDOI{%
\url{https://doi.org/10.1007/978-3-540-69338-3_5}}


\bibitem[\protect\citeauthoryear{Yan and Zhang}{Yan and Zhang}{2008}]%
        {yan2}
\bibfield{author}{\bibinfo{person}{Jun Yan} {and} \bibinfo{person}{Wei Zhang}.}
  \bibinfo{year}{2008}\natexlab{}.
\newblock \showarticletitle{Exploiting Virtual Registers to Reduce Pressure on
  Real Registers}.
\newblock \bibinfo{journal}{{\em ACM Trans. Archit. Code Optim.\/}}
  \bibinfo{volume}{4}, \bibinfo{number}{4}, Article \bibinfo{articleno}{3}
  (\bibinfo{date}{Jan.} \bibinfo{year}{2008}), \bibinfo{numpages}{18}~pages.
\newblock
\showISSN{1544-3566}
\showDOI{%
\url{https://doi.org/10.1145/1328195.1328198}}


\bibitem[\protect\citeauthoryear{You and Chen}{You and Chen}{2015}]%
        {you-vector-register-alloc}
\bibfield{author}{\bibinfo{person}{Yi-Ping You} {and}
  \bibinfo{person}{Szu-Chieh Chen}.} \bibinfo{year}{2015}\natexlab{}.
\newblock \showarticletitle{Vector-aware Register Allocation for GPU Shader
  Processors}. In \bibinfo{booktitle}{{\em Proceedings of the 2015
  International Conference on Compilers, Architecture and Synthesis for
  Embedded Systems}} {\em (\bibinfo{series}{CASES '15})}.
  \bibinfo{publisher}{IEEE Press}, \bibinfo{address}{Piscataway, NJ, USA},
  \bibinfo{pages}{99--108}.
\newblock
\showISBNx{978-1-4673-8320-2}
\showURL{%
\url{http://dl.acm.org.ezproxy.lib.purdue.edu/citation.cfm?id=2830689.2830703}}


\end{thebibliography}

\end{document}